\documentclass[12pt]{article}
\usepackage[utf8]{inputenc} 

\usepackage{draft}
\usepackage{cite} 

\usepackage{hyperref}
\usepackage[margin=1in]{geometry}
\usepackage{setspace}
\usepackage{color}
\usepackage{braket}
\usepackage{titling}
\usepackage{graphicx}
\usepackage[bf,margin=20pt,font={small}]{caption}
\usepackage{subcaption}
\usepackage{youngtab}
\usepackage{epigraph}
\usepackage{tikz}
\usetikzlibrary{math}
\newcommand{\eps}{\epsilon}
\usepackage{amsmath,amssymb,dsfont}
\allowdisplaybreaks

\definecolor{royalblue}{rgb}{0.00000,0.44700,0.74100}
\definecolor{royalorange}{rgb}{0.85000,0.32500,0.09800}
\definecolor{royalyellow}{rgb}{0.92900,0.69400,0.12500}
\definecolor{purple}{rgb}{0.5804, 0.0, 0.82745098}
\definecolor{applegreen}{rgb}{0.55, 0.71, 0.0}
\definecolor{bittersweet}{rgb}{1.0, 0.44, 0.37}

\DeclareMathAlphabet{\mathpzc}{OT1}{pzc}{m}{it}

\usepackage{putex}

\usepackage{mathtools}
\usepackage{siunitx}

\usepackage{pgfplots}
\pgfplotsset{compat=1.10}
\usepgfplotslibrary{fillbetween}
\usetikzlibrary{patterns}

\institution{NYU}{Center for Cosmology and Particle Physics, New York University, New York, NY 10003, USA}

\title{Non-Perturbative Defects in Tensor Models from Melonic Trees}

\authors{Fedor K.~Popov\worksat{\NYU} and Yifan Wang\worksat{\NYU}}

\abstract{
The Klebanov-Tarnopolsky tensor model is a quantum field theory for rank-three tensor scalar fields with  certain quartic potential. The theory possesses an unusual large $N$ limit known as the melonic limit that is strongly coupled yet solvable, producing at large distance a rare example of non-perturbative non-supersymmetric conformal field theory that admits analytic solutions.
We study the dynamics of defects in the tensor model defined by localized magnetic field couplings on a $p$-dimensional subspace in the $d$-dimensional spacetime. While we work with general $p$ and $d$, the physically interesting cases include line defects in $d=2,3$ and surface defects in $d=3$. By identifying a novel large $N$ limit that generalizes the melonic limit in the presence of defects, we prove that the defect one-point function of the scalar field  only receives contributions from a subset of the Feynman diagrams in the shape of melonic trees. These diagrams can be resummed using a closed Schwinger-Dyson equation which enables us to determine non-perturbatively this defect one-point function. At large distance, the solutions we find describe nontrivial conformal defects and we discuss their defect renormalization group (RG) flows. In particular, for line defects, 
we solve the exact RG flow between the trivial and the conformal lines in $d=4-\eps$. We also
compute the exact line defect entropy and verify the $g$-theorem. Furthermore we analyze the defect two-point function of the scalar field  and its decomposition via the operator-product-expansion, providing explicit formulae for one-point functions of bilinear operators and the stress-energy tensor.
}

\date{\today}

\begin{document}
\maketitle 

\tableofcontents

\section{Introduction and Summary}

\subsection{Defects and their conformal limit}
Extended operators known as defects play an important role in the understanding of Quantum Field Theory (QFT). They lead to a vast generalization of observables in QFT beyond the local point-like operators.
Familiar examples include the Wilson loop in gauge theories \cite{Wilson:1974sk} and the Kondo impurity in lattice models \cite{Kondo:1964nea}. More generally, a defect is characterized by its spacetime dimension $p\leq d-1$ or codimension $q=d-p$ and modifies the QFT path integral. On the one hand, living on a submanifold of positive codimension,  a defect can host  an array of novel dynamical phenomena, such as defect renormalization-group (RG) flows \cite{Affleck:1991tk,Affleck:1995ge,Yamaguchi:2002pa,Friedan:2003yc,Gaiotto:2014gha,Jensen:2015swa,Casini:2016fgb,Casini:2018nym,Kobayashi:2018lil,Wang:2021mdq,Cuomo:2021rkm}, phase transitions on the defect \cite{Diehl:1981jgg,Diehl:1983zz,Cardy:1984bb,doi:10.1063/1.447401,Diehl:1996kd,LAW2001159} and defect anomalies \cite{Henningson:1999xi,Graham:1999pm,Schwimmer:2008yh,Cho:2016xjw,Dimofte:2017tpi,Estes:2018tnu,Herzog:2019rke,Bianchi:2019umv,Drukker:2020swu,Herzog:2021hri,Chalabi:2020iie,Drukker:2020dcz, Wang:2020xkc,Wang:2021mdq}, all without altering the physics in the bulk far away from the defect. On the other hand, defects are sensitive to the bulk dynamics and provide an indispensable tool for
detecting generalized higher-form symmetries and anomalies in the bulk \cite{Gaiotto:2014kfa}, which are essential to
elucidate the phase diagram of the bulk QFT \cite{Gaiotto:2017yup}. 

In recent years, the study of defects in QFT have been catalyzed by a combination of numerical and analytical tools that expand on the success in the case of conventional QFT observables such as local correlation functions. This is especially the case for defects in Conformal Field Theory (CFT) whose defect RG fixed points are described by Defect Conformal Field Theory (DCFT). These conformal defects (DCFTs) correspond to universality classes of defects in a given CFT and serve as the starting points to explore the full defect landscape.
Thanks to the extra conformal symmetry
\ie 
SO(p+1,1) \times SO(d-p) \subset SO(d+1,1)\,,
\fe 
preserved by a $p$-dimensional conformal defect, it can be studied numerically by a natural extension of the conformal bootstrap approach
to DCFT  \cite{Liendo:2012hy,Gaiotto:2013nva,Gliozzi:2015qsa,Billo:2016cpy}. 
In cases with supersymmetry,  supersymmetric localization methods have been developed to determine protected defect observables exactly \cite{Pestun:2007rz,Pestun:2009nn,Giombi:2009ek,Kapustin:2009kz,Drukker:2010jp,Kapustin:2012iw,Drukker:2012sr,Kim:2012qf,Assel:2015oxa,Kim:2016qqs,Wang:2020seq}. For non-supersymmetric DCFTs or non-protected observables in supersymmetric DCFTs, it becomes more challenging to obtain analytic results. If the defect has an explicit Lagrangian description, one can always compute the Feynman diagrams as in the standard perturbation theory to determine defect observables. However in practice what one obtains this way is at best an asymptotic series in the small couplings and it is generally difficult to extract non-perturbative results at finite couplings. Several powerful methods have been developed in the last few years to overcome this obstacle using large $N$ \cite{Beccaria:2019dws,Giombi:2020rmc,Herzog:2020lel,Metlitski:2020cqy,Giombi:2021uae,Cuomo:2021kfm}, large charge  \cite{Cuomo:2021cnb,Beccaria:2022bcr,Cuomo:2022xgw,Rodriguez-Gomez:2022gbz}, and integrability \cite{Buhl-Mortensen:2017ind,Grabner:2020nis,Komatsu:2020sup}.\footnote{These methods were developed  building upon previous works that deal with defect-less observables.
The readers can find references to these foundational works in the papers cited here.} In particular, perhaps the simplest interacting scalar field theory, the critical $O(N)$ model (Wilson-Fisher fixed point in $d=4-\epsilon$ dimensions), is known to host a rich family of line defects ($p=1$) connected by nontrivial defect RG flows, and similarly for boundaries ($d-p=1$), which can be studied analytically in the large $N$ limit \cite{Giombi:2020rmc,Herzog:2020lel,Metlitski:2020cqy,Cuomo:2021kfm} (see also related work in the large charge limit at finite $N$ \cite{Cuomo:2021cnb,Cuomo:2022xgw} and previous bootstrap studies at $N=1$ \cite{Gaiotto:2013nva}).

\subsection{Scalar quantum field theory and localized magnetic defects}

Scalar quantum field theory is arguably the most intensively studied class of QFTs. Despite the general obstacle to obtain finite coupling results,  perturbation series in these models can exhibit simplifying features in certain large $N$ limits that make them more tractable. The aforementioned $O(N)$ model is the familiar theory of $N$ scalar fields $\phi_a$ with $a=1,2,\dots N$ and a quartic $O(N)$-invariant potential, also known as the $O(N)$ vector model. The theory has a solvable large $N$ limit that
flows to an interacting but weakly coupled CFT for $d<4$. In particular, at infinite $N$ the scalar fields have classical scaling dimensions (thus free) and the theory has a higher spin symmetry that is weakly broken by anomalous dimensions at $\cO(1/N)$ \cite{Lang:1992zw,Klebanov:2002ja,Maldacena:2011jn,Maldacena:2012sf,Giombi:2016hkj}.\footnote{The weakly broken higher spin symmetry in the $O(N)$ vector model is so constraining that the theory is essentially fixed at the first nontrivial order in the $1/N$ expansion  \cite{Maldacena:2012sf}.}
 There are natural generalizations of the $O(N)$ vector model  in $d\leq 4$ known as tensor models by considering a rank $k$ tensor scalar field  $\phi_{a_1a_2\dots a_k}$ transforming in the fundamental representation of $O(N)^k$, with renormalizable interactions that are invariant under $O(N)^k$.\footnote{See \cite{Klebanov:2018fzb} for a review of the large $N$ tensor models.} These models are in general much more complicated than the vector $O(N)$ case, simply due to the sheer amount of dynamical fields and the proliferation of possible interactions. 
The $k=2$ case is a close analog of the Hermitian matrix model of $N^2$ scalar fields, 
and has a unique single-trace $O(N)^2$-invariant quartic interaction (and one double-trace partner). This matrix theory admits 
a 't Hooft large $N$ limit where the single-trace interaction and planar (fattened) Feynman diagrams dominate \cite{tHooft:1973alw}. It is a much richer case compared to the vector model because already in the leading $N$ limit, there is a nontrivial dependence on the 't Hooft coupling $\lambda_{\rm M}$ (rescaled from the single-trace quartic coupling) and it is a unsolved problem to determine finite $\lambda_{\rm M}$ dependence in matrix models at $d>0$.
Given the incremental complexity from the vector to the matrix $O(N)$ models, one would be surprised to find solvable yet non-trivial tensor models of rank $k\geq 3$. Nonetheless, it has been shown that such a large $N$ limit exists for  tensor models  at $k=3$, where the perturbative expansion is governed by a single $O(N)^3$-invariant quartic coupling known as the tetrahedral interaction and certain (fattened) Feynman diagrams of the melonic type dominate \cite{Klebanov:2016xxf} (generalizing the $d=0$ models in \cite{Gurau:2009tw,Bonzom:2011zz,Carrozza:2015adg} and the $d=1$ fermionic model in \cite{Witten:2016iux}). Thus this limit is commonly referred to as the \textit{melonic limit}. The corresponding tensor model is known as the melonic tensor model which is described by the following action,
\ie 
 S_{\rm TM} = \int d^d x \left[ \frac12 \left(\partial_\mu \phi_{abc}\right)^2 +  {\lambda_{\rm T} \over {4 N^\frac32}} \phi_{abc}\phi_{ab'c'}\phi_{a'bc'}\phi_{a'b'c}  \right]\,.
 \label{TMact}
 \fe 
  As in the matrix large $N$ limit, there is a nontrivial 't Hooft coupling $\lambda_{\rm T}$ (rescaled tetrahedral coupling) in the melonic limit.
 However, unlike in the matrix case, here miraculously, the infinite series in $\lambda_{\rm T}$ can be resummed using an exact Schwinger-Dyson equation, which has lead to the discovery of strongly-coupled scalar CFTs in $d<4$ with order $N^3$ degrees of freedom whose correlation functions can be determined exactly in the leading large $N$ limit \cite{Klebanov:2016xxf}.\footnote{One may wonder what happens for tensor models beyond rank $k=3$. See \cite{Klebanov:2019jup} for a generalization of the solvable melonic limit to the higher rank tensor model at $k\geq 4$, which is governed by a degree $k+1$ ``maximally single-trace'' interaction that generalizes the tetrahedral coupling in \eqref{TMact}.
 }
Thus we see the melonic tensor model serves as an interesting middle ground in the world of scalar QFTs between the vector and matrix models, which is rich yet solvable.\footnote{At the diagrammatic level, one can see explicitly that the melonic diagrams constitute a special subset of the planar diagrams and this makes the resummation possible \cite{Klebanov:2018fzb}.}

Thus far the study of defects in the non-supersymmetric setting is largely restricted to  weakly coupled bulk QFTs, which has already produced many interesting examples with intriguing features \cite{DiPietro:2019hqe,Lauria:2020emq,Behan:2020nsf,DiPietro:2020fya,Behan:2021tcn,Bianchi:2021snj,Cuomo:2021kfm,Cuomo:2022xgw}.\footnote{In particular, even free theories can host nontrivial defects (see for example \cite{DiPietro:2019hqe,Lauria:2020emq,Behan:2020nsf,DiPietro:2020fya,Behan:2021tcn,Bianchi:2021snj}).} Bulk interactions will certainly modify the defect observables in nontrivial ways and it remains a challenge to incorporate their effects
at finite coupling.
Here we will capitalize on the  advantageous features of the melonic tensor models mentioned above to study defects in strongly-coupled scalar QFTs.

More explicitly, we study the following defect in the melonic tensor model  defined by a localized source for \textit{one} component of the scalar field $\phi_{abc}$ on a $p$-dimensional subspace,
\ie 
 S_{\rm DTM}=S_{\rm TM} - \bar J_{abc} \int d^p x_\parallel \, \phi_{abc}(x_\parallel,x_\perp=0) \,,
\label{Mdef}
\fe
where we fix the flavor indices $(abc)$ (no summation).\footnote{It would be interesting to consider more general defect couplings.}  We refer to the coupled system as the \textit{defect tensor model}.
Here we have split the flat spacetime coordinates as $x=(x_\parallel,x_\perp)$ or with indices $x^\m=(x_\parallel^\A,x_\perp^i)$ where $\A=1,2\dots,p$ and $i=p+1,\dots,d$, so that the defect world-volume has coordinates $x_\parallel$ and locates at $x_\perp=0$. This is analogous to the 
pinning field defect (or localized magnetic defect) studied in the context of $(2+1)$-dimensional lattice models \cite{ParisenToldin:2016szc}, where the defect is described by coupling the order parameter to a background magnetic field localized in space and extending in the time direction. The continuum field theory analysis of such defects was carried out in the vector $O(N)$ model in \cite{Cuomo:2021kfm}. We will follow suit and refer to the coupling in \eqref{Mdef} as defining the localized magnetic defect $\cD_p$ in the tensor model. Note that contrary to previous studies of such defects in the vector $O(N)$ model, we do not restrict to line defects (i.e. $p$ is general).\footnote{The physically interesting cases correspond to $d=2$ with $p=1$ and $d=3$ with $p=1,2$ but formally we will work with general $1\leq p<d$. This is possible because in our large $N$ limit \eqref{lim}, the defect problem is reduced to solving a partial differential equation (from the Schwinger-Dyson equation) with a fractional Laplacian in the transverse directions of order $d\over 2$. }

The localized magnetic defect explicitly breaks the bulk $O(N)^3$ global symmetry to an $O(N-1)^3$ subgroup along the defect world-volume. Since the symmetry is unbroken in the bulk, we can re-orient the tensor field in the $O(N)^3$ directions such that the magnetic coupling is with the $\phi_{111}$ component. This defect coupling changes qualitatively 
the correlation functions of local operators built out of $\phi_{abc}$ due to additional Feynman diagrams that can anchor on the defect, in particular 
leading to a nontrivial one-point function $\la \phi_{111}(x)\ra_{\cD_p}$ which otherwise vanishes in the absence of the defect. Note that because of the residual $O(N-1)^3$ symmetry, 
the defect one-point function $\la \phi_{abc}(x)\ra_{\cD_p}$ is zero unless $a=b=c=1$. The main goal here is to study such nontrivial defect correlation functions at the non-perturbative level and we summarize the main results below.

\subsection{Summary of the main results}

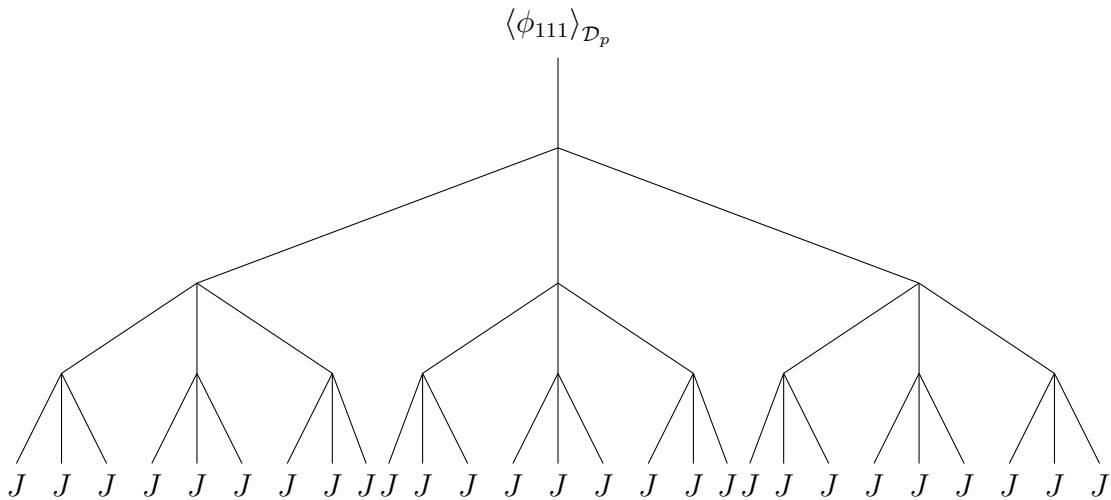
\begin{figure}[!htb]
    \centering
    \begin{tikzpicture}[scale=0.6]
        \draw (0,1)--(-8,-2);
        \draw (0,1)--(0,-2);
        \draw (0,1)--(8,-2);
        \draw (-8,-2)--(-5,-4);
        \draw (-8,-2)--(-8,-4);
        \draw (-8,-2)--(-11,-4);
        \draw (0,-2)--(-3,-4);
        \draw (0,-2)--(0,-4);
        \draw (0,-2)--(3,-4);
        \draw (8,-2)--(5,-4);
        \draw (8,-2)--(8,-4);
        \draw (8,-2)--(11,-4);
        \draw (11,-4)--(11,-6);
        \draw (11,-4)--(12,-6);
        \draw (11,-4)--(10,-6);
        \draw (8,-4)--(7,-6);
        \draw (8,-4)--(8,-6);
        \draw (8,-4)--(9,-6);
        \draw (5,-4)--(6,-6);
        \draw (5,-4)--(5,-6);
        \draw (5,-4)--(4.25,-6);
        \draw (3,-4)--(2,-6);
        \draw (3,-4)--(3,-6);
        \draw (3,-4)--(3.75,-6);
        \draw (0,-4)--(1,-6);
        \draw (0,-4)--(0,-6);
        \draw (0,-4)--(-1,-6);
        \draw (-11,-4)--(-11,-6);
        \draw (-11,-4)--(-12,-6);
        \draw (-11,-4)--(-10,-6);
        \draw (-8,-4)--(-7,-6);
        \draw (-8,-4)--(-8,-6);
        \draw (-8,-4)--(-9,-6);
        \draw (-5,-4)--(-6,-6);
        \draw (-5,-4)--(-5,-6);
        \draw (-5,-4)--(-4.25,-6);
        \draw (-3,-4)--(-2,-6);
        \draw (-3,-4)--(-3,-6);
        \draw (-3,-4)--(-3.75,-6);
        \node[above] at (0,3) {$\braket{\phi_{111}}_{\cD_p}$};
        \draw (0,3)--(0,1);
        \node[below] at (-3.75,-6) {$J$};
        \node[below] at (-3,-6) {$J$};
        \node[below] at (-4.25,-6) {$J$};
        \node[below] at (-5,-6) {$J$};
        \node[below] at (-7,-6) {$J$};
        \node[below] at (-6,-6) {$J$};
        \node[below] at (-8,-6) {$J$};
        \node[below] at (-9,-6) {$J$};
        \node[below] at (-10,-6) {$J$};
        \node[below] at (-11,-6) {$J$};
        \node[below] at (-12,-6) {$J$};
        \node[below] at (3.75,-6) {$J$};
        \node[below] at (3,-6) {$J$};
        \node[below] at (4.25,-6) {$J$};
        \node[below] at (5,-6) {$J$};
        \node[below] at (7,-6) {$J$};
        \node[below] at (6,-6) {$J$};
        \node[below] at (8,-6) {$J$};
        \node[below] at (9,-6) {$J$};
        \node[below] at (10,-6) {$J$};
        \node[below] at (11,-6) {$J$};
        \node[below] at (12,-6) {$J$};
        \node[below] at (0,-6) {$J$};
        \node[below] at (1,-6) {$J$};
        \node[below] at (-1,-6) {$J$};
        \node[below] at (2,-6) {$J$};
        \node[below] at (-2,-6) {$J$};
    \end{tikzpicture}
    \caption{The contribution to the one-point function $\braket{\phi_{111}(x)}_{\cD_p}$ comes only from the tree diagrams, where each edge is the exact propagator (containing nested melons) of the tensor model in the large $N$ limit \eqref{lim}.}
    \label{fig:small_tree}
\end{figure}

While it is straightforward to compute a few of these Feynman diagrams at leading orders in the $\lambda_{\rm T}$ and $\bar J_{111}$ expansion for the one-point function $\la \phi_{111}(x)\ra_{\cD_p}$, 
it may seem a formidable task to obtain the nonperturbative answer which requires summing infinitely many diagrams. The success of the melonic   tensor model without defects motivates us to look for a generalization of the large $N$ limit thereof to the case with defects. Indeed, from a topological argument, we prove in Section~\ref{sec:treeDOM} that in the limit
\ie 
N\to \infty~~{\rm with}~~J\equiv {\bar J_{111}\over N^{3/4}} ~ {\rm fixed} \,,
\label{lim}
\fe
and $\lambda_{\rm T}$ fixed as in the melonic tensor model without defects, the dominant contributions to $\la \phi_{111}(x)\ra_{\cD_p}$ come from \textit{melonic tree} diagrams anchored on the defect with the exact propagator of $\phi_{abc}$ on each edge (see Figure~\ref{fig:small_tree}),
\ie 
\la \phi_{111}(x)\ra_{\cD_p} = N^{3\over 4} f(\lambda_{\rm T},J) + \cO(N^{-{5\over 4}})\,,\quad f(\lambda_{\rm T},J)=({\rm melonic\,trees})\,.
\fe
Furthermore these infinitely many diagrams can be resummed thanks to an exact Schwinger-Dyson equation with the magnetic source, which determines $\la \phi_{111}(x)\ra_{\cD_p}$ (as well as other defect correlation functions) non-perturbatively in $\lambda_{\rm T}$ and $J$ in the leading large $N$ limit \eqref{lim}.

As we will explain in detail in Section~\ref{sec:def1PF}, this compact looking Schwinger-Dyson equation is complicated by the presence of a fractional Laplacian operator $\Delta_\perp^{d\over 4}$ at  $d<4$ in the transverse directions (i.e. $x_\perp$) and it is thus difficult to obtain the closed form solution for $\la \phi_{111}(x)\ra_{\cD_p}$ in general. Nonetheless it is straightforward to solve the Schwinger-Dyson equation in the large distance (IR) limit with a scaling symmetric ansatz for $\la \phi_{111}(x)\ra_{\cD_p}$, which describes the putative conformal regime of the localized magnetic defect. Indeed, the bulk large $N$ tensor model is a nontrivial CFT in the IR limit, where the tensor fields $\phi_{abc}$ are conformal primary operators of dimension $\Delta_\phi={d\over 4}$ \cite{Klebanov:2016xxf,Giombi:2017dtl}.
Consequently, the defect coupling \eqref{Mdef} is a relevant deformation on the $p\geq 1$-dimensional world-volume of a trivial (transparent) defect, and triggers a defect RG flow from the trivial defect to a nontrivial conformal defect. The defect RG and the beta function for the magnetic field $J$ is encoded in the transverse profile of $\la \phi_{111}(x)\ra_{\cD_p}$ which interpolates between the scaling solution we find in the IR and a near UV solution to the Schwinger-Dyson equation close to the defect which we determine via a perturbation theory analysis in $J$ (see Section~\ref{sec:defbf}). In particular, the $J$ expansion can be resummed to produce the exact RG flow for line defects at small $\eps=4-d$ as shown in Section~\ref{sec:linedefRG}.
Moreover, for the special case of codimension-one defects,
we determine the exact solution to the Schwinger-Dyson equation that captures the entire defect RG at finite $\eps=4-d$ (e.g. $d=3$)
in Section~\ref{sec:codimone}. 

For line defects in CFTs, it was recently proven that  the defect RG obeys the $g$-theorem \cite{Cuomo:2021rkm} generalizing previous results in $d=2$ \cite{Friedan:2003yc,Casini:2016fgb}, which states that the $g$-function (its scheme-independent part) or equivalently the defect entropy must decrease monotonically under an RG flow triggered by a relevant deformation on the line defect. In Section~\ref{sec:defgf}, we study the line defect entropy in the tensor model (the $p=1$ case of \eqref{Mdef}). 
We compute the defect entropy exactly in the large $N$ limit in Section~\ref{sec:gfexact} and observe that the $g$-theorem is satisfied for the localized magnetic defect.\footnote{The proof of \cite{Cuomo:2021cnb} uses the locality and unitarity of the defect. More explicitly, the monotonicity of $g$ hinges on the positivity of the two-point function of the trace of the bulk stress-energy tensor along the defect RG flow. Here we find this positivity property holds for the line defect in the melonic tensor model CFT despite the non-unitarity in the bulk operator spectrum (see around \eqref{gfgen}).} For small $\eps=4-d$, we also determine the defect entropy perturbatively in Section~\ref{sec:gfgradient} from explicit Feynman diagrams, verifying our exact result in Section~\ref{sec:gfexact}. 
Furthermore, we check that the gradient formula for line defect RG derived in \cite{Cuomo:2021rkm} holds for our defect tensor model.

We expect the IR limit of the localized magnetic defect in the large $N$ tensor model to be described by a full-fledged non-perturbative DCFT (at least in the leading large $N$ limit). In particular this means that all correlation functions of local operators in the bulk are determined by their bulk operator-product-expansion (OPE) together with the defect one-point functions of general primary operators. The latter can also be determined by diagrammatic techniques in our solvable large $N$ limit \eqref{lim} of the defect tensor model.
With this in mind, in Section~\ref{sec:higherpoint}, we study two-point functions of $\phi_{abc}$ in the DCFT,
and provide formulae for one-point functions of bilinear operators in $\phi_{abc}$, including that of the stress-energy tensor $T_{\m\n}$. We 
leave the more comprehensive analysis to the future.

Having explained the desirable features of the large $N$ tensor model that enable us to solve the localized magnetic defects non-perturbatively, let us insert a word of caution. 
The tensor model defined by the action \eqref{TMact} is not unitary. In particular, the tetrahedral coupling is not positive definite. Instead the theory is described by a complex CFT in the IR limit. Indeed, there exists primary operators of complex scaling dimensions which have been identified in the $\epsilon=4-d$ expansion and to reach the fixed point requires tuning certain couplings to  small but complex values (suppressed in $1\over N$) \cite{Giombi:2017dtl}.
Nonetheless, there is substantial evidence that at least in the leading large $N$ limit, the tensor model \eqref{TMact} is a non-perturbatively well-defined Euclidean CFT \cite{maldacena2016remarks,patashinskii1964second,benedetti2020conformal} and our results here lend further support by incorporating defects.\footnote{Here by non-perturbatively well-defined we mean in the sense of having a spectrum of local operators and OPE coefficients that obey conformal bootstrap equations despite the non-unitarity. 
This is not to be confused with the non-perturbative instability of the conformal solutions discussed in
\cite{Kim:2019upg,benedetti2020melonic,benedetti2021instability}. It is possible to circumvent this instability (and the non-unitarity) by considering the long-ranged version of the tensor model \cite{benedetti2019line} which in particular satisfies the $F$-theorem \cite{benedetti2022f} but no longer has a finite canonical stress energy tensor (as is the case for generalized free fields \cite{Heemskerk:2009pn,El-Showk:2011yvt}).} 
It was recently emphasized in  \cite{Gorbenko:2018ncu}
that complex CFTs are relevant for understanding subtle dynamics of unitary QFTs. For example the presence of complex RG fixed points (in the complexified coupling space) explains the weak first-order phase transitions in statistical models and the walking behavior (slowly-running coupling) of four-dimensional gauge theories below the conformal window. In line with these applications, it would be interesting to understand the implications of the complex melonic CFT and its defects for the unitary tensor model in the neighboring coupling space. Furthermore, there are also generalizations of the large $N$ melonic tensor model that are solvable and manifestly unitary, for example the prismatic theory of \cite{Giombi:2018qgp} in $d\leq 3$ with a positive-definite sextic $O(N)^3$ symmetric interaction, as well as the supersymmetric version in \cite{Popov:2019nja}. It would be interesting to extend our analysis here to defects in those  theories.

\section{Defects in the Large $N$ limit and Tree Dominance}
\label{sec:treeDOM}

For later convenience, we write below the action describing the coupled system of the melonic tensor model with a localized magnetic defect at $x_\perp=0$,
\ie
    S_{\rm DTM} = \int d^d x \left[ \frac12 \left(\partial_\mu \phi_{abc}\right)^2 + \frac{\lambda_{\rm T}}{4 N^\frac32} \phi_{abc}\phi_{ab'c'}\phi_{a'bc'}\phi_{a'b'c} - J N^\frac34 \delta^{d-p}(x_\perp) \phi_{111}  \right]\,.
    \label{eq:action}
\fe
In this section, we investigate the perturbation theory from the above action. From the combinatorial properties of the Feynman diagrams, we will prove that in the large $N$ limit \eqref{lim}, the dominant contributions to the nontrivial defect one-point function $\la \phi_{111}\ra_{\cD_p}$ will come from the melonic trees (see Figure~\ref{fig:small_tree}).

\begin{figure}[!htb]
     \centering
     \begin{tikzpicture}[scale=1.25]
        \draw[red] (1.5,0) circle (1.5);
        \draw[red] (-1,0)--(4,0);
        
        \node[left] at (-1,1.6*sin 10) {$a$};
        \node[left] at (-1,0) {$b$};
        \node[left] at (-1,-1.6*sin 10) {$c$};
        
        \node[right] at (4,1.6*sin 10) {$a$};
        \node[right] at (4,0) {$b$};
        \node[right] at (4,-1.6*sin 10) {$c$};
        
        \draw[green] (1.5+1.6*cos 10,-1.6*sin 10) arc (-10:-170:1.6);
        \draw[green]  (-1,-1.6*sin 10)--(1.5-1.6*cos 10,-1.6*sin 10);
        \draw[green]  (4,-1.6*sin 10)--(1.5+1.6*cos 10,-1.6*sin 10);
        
        \draw[blue] (1.5+1.6*cos 10,1.6*sin 10) arc (10:170:1.6);
        \draw[blue]  (-1,1.6*sin 10)--(1.5-1.6*cos 10,1.6*sin 10);
        \draw[blue]  (4,1.6*sin 10)--(1.5+1.6*cos 10,1.6*sin 10);

        \draw[green] (1.5+1.4*cos 10,1.4*sin 10) arc (10:170:1.4);
        \draw[green] (1.5+1.4*cos 10,1.4*sin 10)-- (1.5-1.4*cos 10,1.4*sin 10);
        
        \draw[blue] (1.5+1.4*cos 10,-1.4*sin 10) arc (-10:-170:1.4);
        \draw[blue] (1.5+1.4*cos 10,-1.4*sin 10)-- (1.5-1.4*cos 10,-1.4*sin 10);

        \draw [-stealth] (4.75,0) -- node [above] {\it  reaping} (6.25,0);
        
        \begin{scope}[xshift=225]
        \draw[blue] (-1,1.6*sin 10)--(1,1.6*sin 10);
        \draw[red] (-1,0)--(1,0);
        \draw[green] (-1,-1.6*sin 10)--(1,-1.6*sin 10);
        
        \node[left] at (-1,1.6*sin 10) {$a$};
        \node[left] at (-1,0) {$b$};
        \node[left] at (-1,-1.6*sin 10) {$c$};
        
        \node[right] at (1,1.6*sin 10) {$a$};
        \node[right] at (1,0) {$b$};
        \node[right] at (1,-1.6*sin 10) {$c$};
        \end{scope}
        
     \end{tikzpicture}
     \caption{Fundamental melon in the tensor model. From the maximal diagram on the left, we   {\it reap} the melon and the overall scaling  stays the same.}
     \label{fig:fund_melon}
 \end{figure}
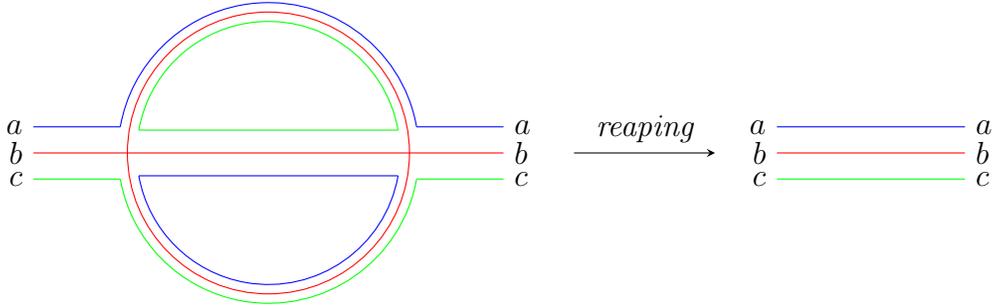

It is well-known that when $J=0$ only specific types of diagrams, known as melonic diagrams (see Figure~\ref{fig:fund_melon}), contribute at the leading order in the melonic large $N$ limit, which allows one to solve the theory (see \cite{Klebanov:2018fzb} for a review). In the case when $J\neq 0$ we should take into account an additional type of diagrams that anchor on the defect world-volume at $x_\perp=0$. Nonetheless, we will show that again only a simple class of diagrams survive in the large $N$ limit \eqref{lim} and the coupled theory \eqref{eq:action} can be solved analytically. 
The proof will be topological (it only depends on the $O(N)^3$ index structures) and therefore  applicable in any dimensions and for any tensor model with the tetrahedral interaction but potentially different field content (e.g. the Gross-Neveu model of fermions $\psi_{abc}$). We also note that the combinatorial result in this section does not depend on the specific form of the source (e.g. localized as in \eqref{eq:action} or not).

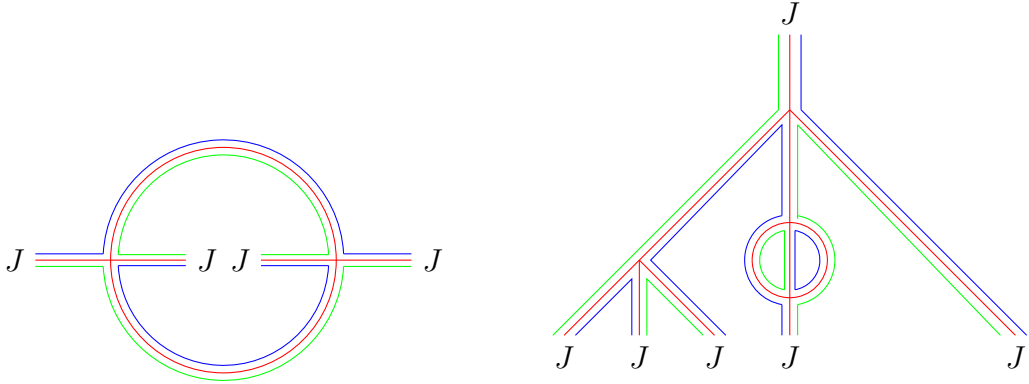
\begin{figure}[!htb]
    \centering
    \begin{tikzpicture}
        \draw[red] (-1,0)--(1,0);
        \draw[red] (1.5,0) circle (1.5);
        \draw[red] (2,0)--(4,0);
        
        \draw[green] (1.5+1.6*cos 3,-1.6*sin 3) arc (-3:-177:1.6);
        \draw[green]  (-1,-1.6*sin 3)--(1.5-1.6*cos 3,-1.6*sin 3);
        \draw[green]  (4,-1.6*sin 3)--(1.5+1.6*cos 3,-1.6*sin 3);
        
        \draw[blue] (1.5+1.6*cos 3,1.6*sin 3) arc (3:177:1.6);
        \draw[blue]  (-1,1.6*sin 3)--(1.5-1.6*cos 3,1.6*sin 3);
        \draw[blue]  (4,1.6*sin 3)--(1.5+1.6*cos 3,1.6*sin 3);

        \draw[green] (1.5+1.4*cos 3,1.4*sin 3) arc (3:177:1.4);
        \draw[green] (1.5+1.4*cos 3,1.4*sin 3)--(2,1.4*sin 3);
        \draw[green] (1.5-1.4*cos 3,1.4*sin 3)--(1,1.4*sin 3);
        
        \draw[blue] (1.5+1.4*cos 3,-1.4*sin 3) arc (-3:-177:1.4);
        \draw[blue] (1.5+1.4*cos 3,-1.4*sin 3)--(2,-1.4*sin 3);
        \draw[blue] (1.5-1.4*cos 3,-1.4*sin 3)--(1,-1.4*sin 3);

        \draw[left] node at (-1,0) {$J$};
        \draw[left] node at (2,0) {$J$};
        \draw[right] node at (4,0) {$J$};
        \draw[right] node at (1,0) {$J$};

        \begin{scope}[xshift=200]
            \draw[red] (-1,-1)--(0,0);
            \draw[red] (0,-1)--(0,0);
            \draw[red] (1,-1)--(0,0);
            \draw[red] (0,0)--(2,2);
            \draw[red] (2,2)--(2,-1);
            \draw[red] (2,2)--(5,-1);
            \draw[red] (2,2)--(2,3);
            \draw[red] (2,0) circle (0.5);

            \draw[blue] (2.15,3)--(2.15,2);
            \draw[blue] (-0.85,-1)--(-0.1,-0.25)--(-0.1,-1);
            \draw [blue]  (2.15,2)--(5.15,-1);
            \draw [blue] (1.15,-1)--(0.15,0)--(2.0+0.6*cos 100,1.8)--(2.0+0.6*cos 100,0.6*sin 100);
            \draw [blue] (2.0+0.6*cos 100,0.6*sin 100) arc (100:260:0.6);
            \draw [blue] (2.0+0.4*cos 80,0.4*sin 80) arc (80:-80:0.4);
            \draw [blue] (2.0+0.6*cos 100,-0.6*sin 100)--(2.0+0.6*cos 100,-1);
            \draw [blue] (2.0+0.4*cos 80,0.4*sin 80)--(2.0+0.4*cos 80,-0.4*sin 80);

            \draw[green] (1.85,3)--(1.85,2);
            \draw[green] (1.85,2)--(-1.15,-1);
            \draw[green] (0.85,-1)--(0.1,-0.25)--(0.1,-1); 
            \draw [green] (2+0.4*cos 100,0.4*sin 100) arc (100:260:0.4);
            \draw [green] (2+0.6*cos 80,0.6*sin 80) arc (80:-80:0.6);
            \draw [green] (2+0.6*cos 80,-0.6*sin 80)--(2+0.6*cos 80,-1);
            \draw [green] (2+0.4*cos 100,0.4*sin 100)--(2+0.4*cos 100,-0.4*sin 100);
            \draw [green] (2+0.6*cos 80,1.8)--(2+0.6*cos 80,0.55);
            \draw [green] (2+0.6*cos 80,1.8)--(4.8,-1);

            \draw[below] node at (-1,-1) {$J$};
            \draw[below] node at (0,-1) {$J$};
            \draw[below] node at (1,-1) {$J$};
            \draw[below] node at (2,-1) {$J$};
            \draw[below] node at (5,-1) {$J$};
            \draw[above] node at (2,3) {$J$};

        \end{scope}
    \end{tikzpicture}
    \caption{On the left is an example of a planar diagram that is subleading in the large $N$ limit \eqref{lim}. On the right is an example of the melonic tree diagrams that dominate in this limit.}
    \label{fig:diags}
\end{figure}

 Let us consider an arbitrary Feynman diagram from the action \eqref{eq:action}, that contains $V$ vertices with tetrahedral interactions, $S>0$ sources and $E$ propagators. Some examples of these diagrams are depicted in Figure~\ref{fig:diags}. The task is to find the $N$ scalings of the vertices and sources so as to have a smooth large $N$ limit (which will be \eqref{lim} as promised) and to identify the subset of the diagrams that dominate in this limit.
 
  We follow the standard strategy that is employed in the tensor models without defects (see \cite{Klebanov:2018fzb} for a review). We make explicit the index structure for each Feynman diagram by resolved (stranded) graphs where each edge consists of three colored strands representing the propagation of the three indices of $\phi_{abc}$ and the strands are joined one-by-one at the vertices preserving the color. We denote the number of index loops for each color by $F_i$ and the total number of loops by $F=F_1+F_2+F_3$.
  The $N$ and coupling dependence of the contribution from such a diagram is 
  \ie 
  N^F \left(\lambda_{\rm T}\over N^{3/2}\right)^V\left( J N^{3/4}\right)^S = N^{F-{3\over 2}  V+{3\over 4}S} \lambda_{\rm T}^V J^S\,.
  \label{eachdiag}
  \fe
  If we erase one of the colors from our diagram, we obtain a ribbon graph (fat graph), of the same type that appears in the matrix model. Consequently we can assign a genus $g_{ij}$ for each ribbon graph where $i,j$ denote  the colors of the remaining strands and $F_{ij}=F_i+F_j$ counts the number of index loops in the ribbon graph. Furthermore, since we are dealing with open ribbon graphs (due to the $J$ sources), the corresponding surface has $n_{ij}$ boundary components. For instance, for the left diagram in the Figure~\ref{fig:diags} we have $n_{ij}=2$ and for the right diagram we have $n_{ij}=1$. 
  We then arrive at the following combinatorial relation from the Euler characteristic,
\begin{gather}
    (2-n_{ij}) - 2 g_{ij} = V+S - E + F_{ij} \quad {\rm with}~g_{ij} \geq 0\,,~n_{ij}\geq 1\,,
    \label{eq:Euler}
\end{gather}
Since each propagator terminates either at a vertex or at the source term,  we also have the following obvious relation 
\begin{gather}
    4V+S=2E\,. \label{eq:VK}
\end{gather}
Combining equations \eqref{eq:Euler} and \eqref{eq:VK} we obtain, for a connected graph,
\begin{gather}
3-\frac{1}{2}n - g + \frac32 V - \frac34 S = F ~\Rightarrow~      F-{3\over 2}  V+{3\over 4}S \leq {3\over 2} \,, \label{eq:Fbound}
\end{gather}
where we have used $g\equiv g_{12}+g_{23}+g_{13} \geq 0$ and $n\equiv n_{12}+n_{23}+n_{13} \geq 3$. Comparing with \eqref{eachdiag}, we thus find that to achieve a smooth large $N$ limit we should keep $\lambda_{\rm T}$ and $J$ fixed as promised, and the dominant contributions can scale at most as $N^{3\over 2}$ for a connected graph.

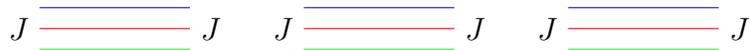
\begin{figure}[!htb]
    \centering
    \begin{tikzpicture}
        \draw[blue] (-1,1.6*sin 10)--(1,1.6*sin 10);
        \draw[red] (-1,0)--(1,0);
        \draw[green] (-1,-1.6*sin 10)--(1,-1.6*sin 10);
        \node[left] at (-1,0) {$J$};
        \node[right] at (1,0) {$J$};
        \begin{scope}[xshift=100]
        \draw[blue] (-1,1.6*sin 10)--(1,1.6*sin 10);
        \draw[red] (-1,0)--(1,0);
        \draw[green] (-1,-1.6*sin 10)--(1,-1.6*sin 10);
        \node[left] at (-1,0) {$J$};
        \node[right] at (1,0) {$J$};
        \end{scope}
        \begin{scope}[xshift=200]
        \draw[blue] (-1,1.6*sin 10)--(1,1.6*sin 10);
        \draw[red] (-1,0)--(1,0);
        \draw[green] (-1,-1.6*sin 10)--(1,-1.6*sin 10);
        \node[left] at (-1,0) {$J$};
        \node[right] at (1,0) {$J$};
        \end{scope}
    \end{tikzpicture}
    \caption{An example of an empty graph, from a collection of disjoint empty trees that does not contain any interaction vertices but contribute to the leading order in the large $N$ limit.}
    \label{fig:emptytree}
\end{figure}

Now let us identify the \textit{maximal graphs}, namely graphs whose connected components saturate the inequality \eqref{eq:Fbound}. We focus on a connected maximal graph, which is achieved if and only if  $g = 0$ and $n=3$. In other words, all ribbon graphs obtained from erasing one color are planar and each has a single boundary.
For example, the left diagram of Figure~\ref{fig:diags} is non-maximal because $n_{bg}=2$ for the blue and green colors, while $n_{br}=n_{gr}=1$ for the other choices of colors (and so $g=0$ but $n=4$).
For a resolved graph, a strand either forms a closed loop or connects two sources. 
For strands that pass through exactly $m$ vertices, we denote the number of index loops by $P_m$ and count those ending on sources by $L_m$. For a maximal graph, we have the following immediate relation,\footnote{We focus on connected graphs here so that any loop must pass through at least one vertex (i.e. $P_0=0$).}
 \ie 
F= \sum_{m\geq 1} P_m = {3\over 2}+{3\over 2} V-{3\over 4}S\,.
\label{comb1}
 \fe
 Since each tetrahedral vertex is passed through six times and each source has three outgoing strands, we also have 
 \ie 
 \sum_{m\geq 1} m (P_m + L_m) = 6 V\,,\quad  \sum_{m\geq 0} L_m = \frac32 S\,.
 \label{comb2}
 \fe 
 As explained in \cite{Klebanov:2018fzb}, for the ribbon graphs (after erasing a color) to be planar, one must have $P_1=P_3=0$. Then putting together \eqref{comb1} and \eqref{comb2}, we find
 \ie 
 2 P_2 + L_1 +2L_0 = 6 + \sum_{m\geq 5}(m-4) P_m +\sum_{m \geq 3} (m-2) L_m \geq 6\,.   
 \fe 
 If $L_0>0$, the only possible connected graph is simply two sources connected by an edge and thus $L_0=3$ (and $P_m=L_{m>0}=0$). This is clearly a maximal graph, which we define as the connected \textit{empty graph}.  A general empty graph is defined as a disjoint union of such graphs. They are maximal graphs which contain  no loops or vertices 
(see Figure~\ref{fig:emptytree}).
More nontrivial connected maximal graphs arise when $L_0=0$, in which case the graph must either has a loop that passes through exactly two vertices ($P_2>0$) or a strand that goes through exactly one vertex and connects two sources ($L_1>0$).

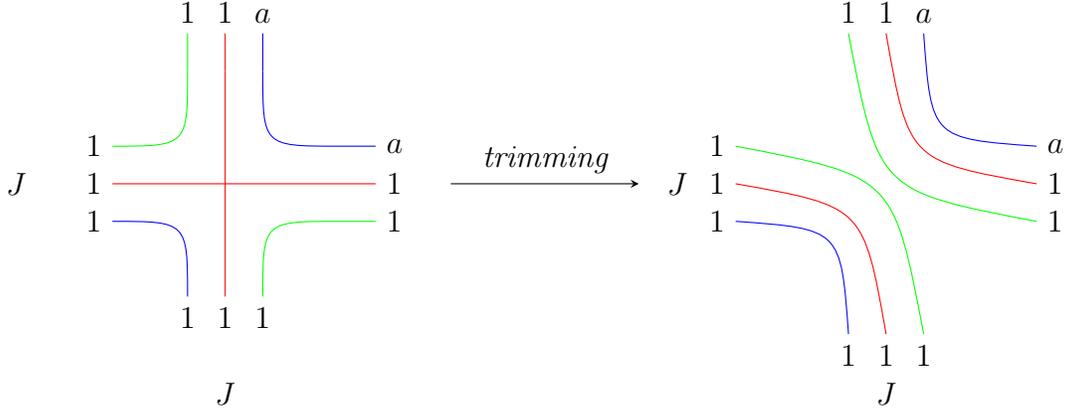
\begin{figure}[!htb]
    \centering
    \begin{tikzpicture}
        \draw[green] (-1.5,0.5)..controls (-0.5,0.5)..(-0.5,1.5);
        \draw[blue] (-1.5,-0.5).. controls (-0.5,-0.5)..(-0.5,-1.5);
        \draw[blue] (1.5,0.5).. controls (0.5,0.5)..(0.5,1.5);
        \draw[green] (1.5,-0.5).. controls (0.5,-0.5)..(0.5,-1.5);

        \draw[red] (-1.5,0)--(1.5,0);
        \draw[red] (0,-1.5)--(0,1.5);
        
        \draw[red] (1.5,0) -- (2,0);
        \draw[blue] (1.5,0.5) -- (2,0.5);
        \draw[green] (1.5,-0.5) -- (2,-0.5);
        
        \draw[red] (0,1.5) -- (0,2);
        \draw[blue] (0.5,1.5) -- (0.5,2);
        \draw[green] (-0.5,1.5) -- (-0.5,2);
        
         \draw[below] node at (0,-2.5) {$J$};
        \draw[left] node at (-2.5,0) {$J$};
        \node[left] at (-1.5,0) {$1$};
        \node[left] at (-1.5,-0.5) {$1$};
        \node[left] at (-1.5,0.5) {$1$};
        
        \node[below] at (0,-1.5) {$1$};
        \node[below] at (-0.5,-1.5) {$1$};
        \node[below] at (0.5,-1.5) {$1$};
        
        \node[right] at (2,0) {$1$};
        \node[right] at (2,-0.5) {$1$};
        \node[right] at (2,0.5) {$a$};
        
        \node[above] at (0,2) {$1$};
        \node[above] at (-0.5,2) {$1$};
        \node[above] at (0.5,2) {$a$};
        
        \draw [-stealth] (3,0) -- node [above] {\it trimming} (5.5,0);
        
        \begin{scope}[xshift=250]

        \draw[blue] (-2,-0.5) .. controls (-0.6,-0.6) .. (-0.5,-2);
        \draw[red] (-2,0) .. controls (-0.3,-0.3) .. (0,-2);
        \draw[green] (-2,0.5) .. controls (0.1,0.1) .. (0.5,-2);

        \draw[blue] (2,0.5) .. controls (0.6,0.6) .. (0.5,2);
        \draw[red] (2,0) .. controls (0.3,0.3) .. (0,2);
        \draw[green] (2,-0.5) .. controls (-0.1,-0.1) .. (-0.5,2);
        
        \node[right] at (2,0) {$1$};
        \node[right] at (2,-0.5) {$1$};
        \node[right] at (2,0.5) {$a$};
        
        \node[above] at (0,2) {$1$};
        \node[above] at (-0.5,2) {$1$};
        \node[above] at (0.5,2) {$a$};
        
        \draw[below] node at (0,-2.5) {$J$};
        \draw[left] node at (-2.5,0) {$J$};
        \node[left] at (-2,0) {$1$};
        \node[left] at (-2,-0.5) {$1$};
        \node[left] at (-2,0.5) {$1$};
        
        \node[below] at (0,-2) {$1$};
        \node[below] at (-0.5,-2) {$1$};
        \node[below] at (0.5,-2) {$1$};
        \end{scope}
    \end{tikzpicture}
    \caption{If we have a vertex such that it is connected to two sources $J$ we can perform the trimming surgery and remove these two sources without spoiling maximality. }
    \label{fig:trimming}
\end{figure}

Below we prove that the maximal graph must be a melonic tree recursively. The strategy is to introduce two \textit{surgery} operations on a resolved graph,  \textit{trimming} and \textit{reaping}, as in Figure~\ref{fig:trimming} and Figure~\ref{fig:fund_melon} respectively. Intuitively, trimming cuts off branches of the graph that straddle neighboring sources, and reaping removes melonic sub-graphs, replacing them with bare branches.
Importantly, these surgery operations reduce the graph by removing vertices and loops while preserving maximality.\footnote{In particular, the $N$ dependence is preserved if the reduced graph remains connected.}   We will argue below that any maximal graph can be reduced by these two operations to an empty (disconnected) graph as in Figure~\ref{fig:emptytree}. Undoing the surgeries, we then establish that all maximal graphs are (possibly disconnected) trees connecting the sources with melons on the tree branches (edges).

We start with the case of $L_1>0$. This requires a vertex with a pair of adjacent branches that anchor on the neighbouring sources. It is easy to see that we can perform a trimming surgery as in Figure~\ref{fig:trimming}, removing this vertex and producing a disconnected empty sub-graph. We can repeat this surgery operation until all such vertices are removed. 

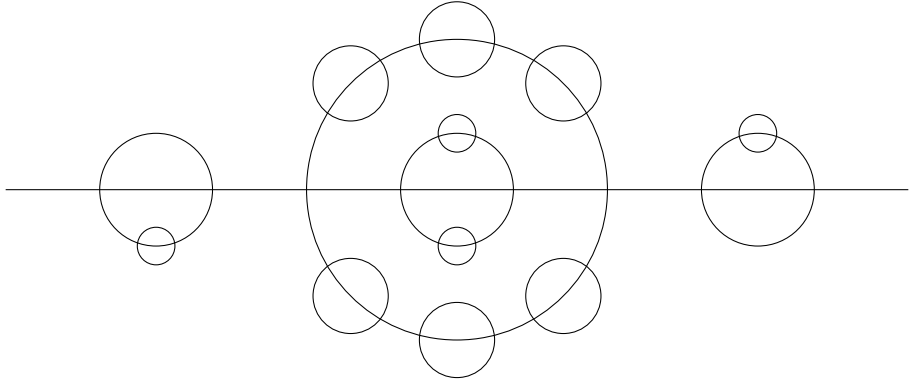
\begin{figure}[!htb]
    \centering
    \begin{tikzpicture}
        \draw (-6,0)--(6,0);
        \draw (0,0) circle (2);
        
        \draw (0,0) circle (0.75);
        \draw (0,2) circle (0.5);
        \draw (1.414,1.414) circle (0.5);
        \draw (-1.414,-1.414) circle (0.5);
        \draw (1.414,-1.414) circle (0.5);
        \draw (-1.414,1.414) circle (0.5);
        \draw (4,0) circle (0.75);
        \draw (-4,0) circle (0.75);
        \draw (0,-2) circle (0.5);
        \draw (-4,-0.75) circle (0.25);
        \draw (4,0.75) circle (0.25);
        \draw (0,0.75) circle (0.25);
        \draw (0,-0.75) circle (0.25);
    \end{tikzpicture}
    \caption{Nested melons on an edge.}
    \label{fig:nestmelon}
\end{figure}

In the case $P_2>0$, the situation is similar to the usual defect-less tensor model (i.e. graphs with $S=0$). Here we have a subgraph that contains a pair of vertices connected by one loop. This loop separates the plane into two disconnected regions (see for example the left diagrams of Figure~\ref{fig:diags} and Figure~\ref{fig:fund_melon}). Since $n_{ij}=1$ (for the ribbon graph after erasing one color) one of these regions does not contain any sources (the left diagram of Figure~\ref{fig:diags} does not satisfy this condition and therefore is not maximal). In that region we can use the standard proof of melonic dominance   \cite{Bonzom:2011zz,Klebanov:2016xxf,Klebanov:2018fzb}  and conclude that it is an edge nested with melons, as in Figure~\ref{fig:nestmelon}. This means that the entire subgraph we start with is also a edge (branch) with melons.
Therefore we can apply the reaping surgery to remove all melons as in Figure~\ref{fig:fund_melon}, leaving behind a bare branch.

Combining with trimming operation above, we can then reduce all maximal graphs to an empty graph. Tracing backward, we establish that all maximal graphs are melonic trees (e.g. the right diagram of Figure~\ref{fig:diags}).

The tree dominance in the large $N$ limit \eqref{lim} dramatically simplifies the diagrammatics of the defect tensor model. In the following sections, we will exploit this feature to compute various observables for the localized magnetic defects \eqref{Mdef} in the tensor model. 
 
Before ending this section, let us comment on how the melonic tree dominance is affected if we consider localized magnetic defects in other generalizations of the tensor model, or more general types of defect couplings than those in \eqref{eq:action}.

There are generalizations of the tensor model with a single $O(N)$ symmetry group where the field $\phi_{abc}$ transforms in a non-trivial rank-three $O(N)$ representation \cite{klebanov2017large,carrozza2018large} which admit defects of the form \eqref{eq:action}. Even though we do not have a complete proof, we still expect that in these models, melonic tree diagrams dominate in the large $N$ limit. Meanwhile, in  other tensor-like models, such as the Gurau-Witten model \cite{witten2019syk} or SYK-like models \cite{liu2019d}, the non-trivial melonic tree diagrams that will play an important role here for defects would be suppressed in the large $N$ limit. Nevertheless, it is curious to note that in the context of the SYK model, the melonic trees appear  in the study of operator growth   
\cite{roberts2018operator}, where they  encode  the evolution of the complexity of the initial state with time.

One generalization of the defect coupling in \eqref{eq:action} within the melonic tensor model is to turn on the $O(N)^3$ singlet operator $\phi_{abc}^2$ on the defect world-volume. This is a localized mass deformation that does not break any global symmetry. Instead of melonic trees, the large $N$ contributions for defect observables (e.g. the one-point function of $\phi_{abc}^2$) now come from melonic ladder diagrams. The corresponding Schwinger-Dyson equation (see Section~\ref{sec:defSDE}) becomes more complicated.

\section{The Schwinger-Dyson Equation and Defect One-point Functions}
\label{sec:def1PF}

\subsection{Defect Schwinger-Dyson equation} 
\label{sec:defSDE}

We proceed to compute the partition function of the defect tensor model \eqref{eq:action} as a function of the external sources $J$ in the large $N$ limit \eqref{lim}. For the moment we will keep the position dependence of the source general (i.e. do not assume that it takes the specific localized form in \eqref{eq:action}).

We first note that all melon contributions (see Figure~\ref{fig:nestmelon}) to the \textit{exact} propagator $G(p)$ of  $\phi_{abc}$ can be resummed   
by the means of a Schwinger-Dyson equation that is closed in the melonic limit \cite{Klebanov:2016xxf},
\begin{gather}\braket{\phi_{abc}(p)\phi_{a'b'c}(-p)} = \delta_{aa'}\delta_{bb'}\delta_{cc'} G(p)\,, \notag\\
G^{-1}(p) = p^2 - \lambda_{\rm T}^2 \int \frac{d^d k}{(2\pi)^d} \frac{d^d q}{(2\pi)^d} G(p-k-q)G(k)G(q)\,, \label{eq:DSprop}
\end{gather}
where we are working with the momentum space and $p^2$ is the inverse of the bare propagator. In the IR limit (at large distances) of this massless theory, we expect an emergent conformal symmetry  and thus a nontrivial solution to \eqref{eq:DSprop} with definite scaling behavior given by,\footnote{Note that in our convention, the prefactor of \eqref{eq:Gexact} explicitly depends on the bare coupling $\lambda_{\rm T}$, but this dependence cancels out in all physical observables in the DCFT, such as the defect entropy, dimensions of operators and (normalized) OPE coefficients.} 
\begin{gather}
    G(x) = \frac{1}{\sqrt{\lambda_{\rm T}}}\frac{A_{d}}{x^\frac{d}{2}} \,,\quad
    A_d=  \left(\frac{d\Gamma\left(\frac{3d}{4}\right)}{ 4\pi^d\Gamma\left(1-\frac{d}{4}\right)}\right)^\frac14 \,, \notag\\
 G^{-1}(x) = \sqrt{\lambda_{\rm T}} B_d
 (-\Delta)^\frac{d}{4}     \,,\quad
    B_d= \left(\frac{4\Gamma\left(1-\frac{d}{4}\right)}{ d(4\pi)^d\Gamma\left(\frac{3d}{4}\right)}\right)^\frac14
   \,, 
  \label{eq:Gexact}
\end{gather}
where $\Delta$ is the Laplacian and we have used the following relation
\begin{gather}
\int d^d x\, \frac{e^{i k x}}{|x|^{a}} =  \frac{ f(d,a) }{|k|^{d-a}}\,,\quad 
 f(d,a)= \frac{\pi^\frac{d}{2} \Gamma\left(\frac{d-a}{2}\right)}{2^{a-d} \Gamma\left(\frac{a}{2}\right)} \frac{1}{|k|^{d-a}}\,.
\label{eq:VeryUsefulFormula}
\end{gather}

Now in the presence of the sources $J$, we need to resum all possible melon contributions to a given (bare) tree diagram (for instance the right diagram of Figure~\ref{fig:diags} adds a one-melon contribution to the bare tree diagram). This boils down to replacing all edges  with the exact propagator \eqref{eq:Gexact}. The partition function for the tensor model $Z[J]$ as a function of the external source $J$ can then be computed by recursively generating all possible tree diagrams. For our purpose, we consider the one-point function 
\ie 
F(x)\equiv \frac{1}{N^\frac34}\braket{\phi_{111}(x)}\,,
\label{Fdef}
\fe 
as a function of $J$, which satisfies the following Schwinger-Dyson equation
\begin{gather}
    F(x) = \int d^d y\, G(x-y)\,\left[ J(y) - \lambda_{\rm T} F^3(y)\right]\,.
    \label{SDEintegral}
\end{gather}
The above equation, which is a consequence of the tree dominance in the large $N$ limit, is reminiscent of the expression for a classical defect in the classical tensor model. The only difference is that the full propagator $G(x-y)$ would be replaced by the bare propagator $G_0(x-y)$.
In other words, all quantum corrections for the  defect tensor model in the large $N$ limit is entirely captured by the propagator of $\phi_{abc}$.
Partly this happens because the vertex function does not receive any correction in the large $N$ limit. We emphasize that in the case of critical $O(N)$ model or matrix model, there are other contributions to the equation above coming from the quantum corrections to the vertex functions. 

Up to this point in the section, we have not assumed a specific form for the external source $J(x)$ for the scalar $\phi_{111}(x)$. For the rest of the paper, we will restrict the source to be  localized on a subspace of dimension $p$ as in \eqref{eq:action} which defines the localized magnetic defect in the tensor model (i.e. $J(x) \to J \delta^{d-p}(x_\perp)$). The goal for the rest of the paper is to solve the defect Schwinger-Dyson equation and determine defect observables.

\subsection{Conformal defect at large distance}
\label{sec:confdf}
For $d<4$ the tensor model flows to a non-trivial melonic CFT at large distance, which is described by the conformal solution  \eqref{eq:Gexact} to the bulk Schwinger-Dyson equation \eqref{eq:DSprop} in the large distance limit.

Using the translation symmetry in the $x_\parallel$ directions for the defect, $F(x)$ clearly only depends on $x_\perp$. The defect Schwinger-Dyson equation \eqref{SDEintegral} for $F(x_\perp)$ as defined in \eqref{Fdef} can then be put into the following form,\footnote{There are two characteristic scales in the coupled system \eqref{eq:action}: one set by the UV cutoff in the bulk, and the other set by the bare defect coupling.
We emphasize that we are studying defects in the melonic CFT where the UV cutoff has been sent to infinity. For that purpose, we are using the conformal solution \eqref{eq:Gexact} instead of the exact propagator that follows from the bulk Schwinger-Dyson equation \eqref{eq:DSprop}. 
}
\footnote{Here and in the rest of the paper unless otherwise specified, $\lambda_{\rm T}$ is the renormalized dimensionless coupling,  $\phi_{abc}$ is the renormalized operator that is a conformal primary and correspondingly $J$ is the renormalized source.}
\begin{gather}
 \sqrt{\lambda_{\rm T}}  B_d (-\Delta_\perp)^\frac{d}{4} F + \lambda_{\rm T} F^3 = J \delta^{d-p}(x_\perp)\,,
 \label{SDEdiff}
\end{gather}
which now involves  a fractional Laplacian  which is naturally defined by Fourier transformation from the momentum space, as in
\ie 
(-\Delta)^{\A} f(x)\equiv \int {d^{d} k\over (2\pi)^d} \int d^d y \,e^{ik(x-y)}|k|^{2\A}  f(y)\,,
\fe
for Laplacian $\Delta$ on the entire spacetime and similarly for the transverse Laplacian $\Delta_\perp$. The constant $B_d$ is defined in \eqref{eq:Gexact}.

The magnetic defect induces a nontrivial one-point function for local operators, as for the classical case. However here $\braket{\phi_{111}}$ has a different spatial dependence due to the anomalous dimension of the operator $\Delta_\phi={d\over 4}$ in the melonic CFT. While the Schwinger-Dyson equation \eqref{SDEdiff} is difficult to solve, its IR limit simplifies and we expect its solution to describe a nontrivial DCFT.   As is well-known, the one-point function of a bulk primary in DCFT is completely fixed up to an overall constant \cite{Billo:2016cpy}. Here we have 
\begin{gather}
\label{scant}
\braket{\phi_{111}(x)} =  {N^\frac34\over \lambda_{\rm T}^{1\over 4}}\frac{C_{d,p}}{|x_\perp|^\frac{d}{4}}\,.
\end{gather}
To determine the coefficient $C_{d,p}$, it is convenient to work with the integral form of the Schwinger-Dyson equation \eqref{SDEintegral}, which simplifies in the IR limit to the following,
\ie 
F(x)=-\lambda_{\rm T} \int d^d y G(x-y) F^3(y)\,.
\label{SDEconf}
\fe
This leads to the following integral 
\begin{gather}
\frac{C_{d,q}}{|x_\perp|^\frac{d}{4}}=- \lambda_{\rm T} C_{d,q}^3 A_d \int d^{d-p} y_\perp d^{p} y_\parallel 
\frac{1}{(|x_\perp-y_\perp|^2+|y_\parallel|^2)^{d\over 4} |y_\perp|^\frac{3d}{4}}\,,
\end{gather}
which can be evaluated by standard methods and gives,
\ie 
C^2_{d,p}={\Gamma \left(3d\over 8\right) \Gamma \left(3d-4p\over 8\right) (-\Gamma \left(-{d\over 4}\right))^{1\over 4}
\over \pi^{d\over 4} \Gamma \left( d\over 8\right) (-\Gamma \left(d-4p\over 8\right)) \Gamma\left(3d\over 4\right)^{1\over 4}}
\,.
\fe
Note that the RHS above is manifestly positive for $1\leq p<d$ and $p+1\leq d<4$. It may seem that information of the source $J$ has completely disappeared in the IR limit. However physically we expect the sign for the one-point function to be fixed by $\sgn(J)$ of the localized magnetic field. Before explaining how we derive this sign, let us state the result.
We find the one-point function for the DCFT to be
\ie 
\la \phi_{111}(x)\ra_{\rm normalized}
= \sgn(J)\left( {  
  \Gamma \left( {3d\over 8} \right)
\Gamma \left( {3d -4p \over 8}  \right) 
 (-\Gamma \left( -{d\over 4}\right))^{1\over 2}
\over 
\Gamma\left (d\over 8\right)   \left(-\Gamma \left( {d -4p\over 8}  \right) \right)\Gamma \left( 3d\over 4\right)^{1\over 2}  
} \right)^{1\over 2}
{N^{3\over 4}\over |x_\perp|^{d\over 4}}\,,
\label{norm1PF}
\fe
where we have normalized $\phi_{111}$ by its two point function. As we will see, the overall sign is fixed by considering the case $p=d-1$ in Section~\ref{sec:codimone} and for line defects in Section~\ref{sec:linedefRG}.

We emphasize that the simplified Schwinger-Dyson equation \eqref{SDEintegral}   applies because the operator $\phi_{111}(x_\parallel,x_\perp=0)$ is a relevant deformation on the worldvolume of the trivial $p\geq 1$-dimensional defect, so that the part of \eqref{SDEintegral} that depends explicitly on the source $J$ becomes unimportant in the IR limit. Later we will study in detail the defect RG flow that interpolates between the UV coupling as in \eqref{eq:action} and the IR DCFT. Here we comment that by consistency if the defect coupling were irrelevant, the scaling solution \eqref{scant} would not apply. For example in the defect tensor model action \eqref{eq:action} if we have instead a coupling to conformal descendants of $\phi_{111}$,
\ie 
  - J_{\rm irrel} N^{3\over 4} \int d^p x_\parallel\, (\Delta_\perp)^n \phi_{111}(x_\parallel,x_\perp=0)\,,
\fe
which is irrelevant for sufficiently large $n$, the combinatorics is unaffected and thus the tree dominance proved in Section~\ref{sec:treeDOM} holds, and consequently we arrive at a similar defect Schwinger-Dyson equation as in \eqref{SDEintegral}
\ie 
F(x)=J_{\rm irrel}  \Delta_\perp^n G (x_\perp) - \lambda_{\rm T} \int d^{d-p} y_\perp G_\perp(x_\perp-y_\perp) F^3(y)\,,
\label{SDEirr}
\fe
where 
\ie 
 G_\perp(x) \equiv \int d^p x_\parallel {G}(x_\perp, x_\parallel)= 
\pi ^{p/2}  A_d   \frac{\Gamma \left(   d-2 p\over 4\right)}{\Gamma \left(\frac{d}{4}\right)} \frac{1}{x_\perp^{\frac{d}{2}-p}}\,.
\fe
In this case, it is easy to see that $F(x)$ is dominated at large $x_\perp$ by the classical contribution from the source term in \eqref{SDEirr} as expected,
\ie 
 F(x) \to    \frac{J_{\rm irrel}}{x_\perp^{\frac{d}{2}+2n-p}}\,,
\fe
up to a constant that is independent of the couplings. This faster fall-off (than in \eqref{scant}) is to be interpreted as a vanishing one-point function in the IR limit where the defect flows to the trivial one.

Finally we mention in passing that the conformal defect one-point function 
\eqref{scant} in the rank-three tensor model has a simple generalization for general rank-$k$ tensor models \cite{Klebanov:2019jup}. In this case, the coefficient $C_{d,p}^{(k)}$ satisfies,
\ie 
    \left(C^{(k)}_{d,p}\right)^{k-2} 
    =&  - \frac{\pi ^{\frac{d}{k}-\frac{d}{2}} \Gamma \left(\frac{d}{k}\right) \Gamma \left(\frac{d
   (k-1)}{2 k}\right) \Gamma \left(\frac{1}{2}
   \left(-\frac{d}{k}+d-p\right)\right)}{\Gamma \left(\frac{d}{2 k}\right) \Gamma
   \left(\frac{d (k-2)}{2 k}\right) \Gamma \left(\frac{d-k p}{2 k}\right)}
   \left(-\frac{\Gamma \left(d \left(\frac{1}{k}-\frac{1}{2}\right)\right) \Gamma
   \left(\frac{d (k-2)}{2 k}\right)}{\Gamma \left(\frac{d}{k}\right) \Gamma \left(\frac{d
   (k-1)}{k}\right)}\right)^{1\over k}\,.
\fe

\subsection{Analytic defect one-point functions for codimension-one defects}
\label{sec:codimone}
While in general it is difficult to solve for the complete profile of the one-point function $\la\phi_{111}\ra_{\cD_{d-1}}$ from the Schwinger-Dyson equation \eqref{SDEdiff} (or \eqref{SDEintegral}), the equation simplifies for the codimension one defects (i.e. $p=d-1$) and we will determine the exact solution here.

As a warm-up, let us consider the differential equation that describes a classical point-like (i.e. $p=0$) defect at $d=1$, 
\begin{gather}
    \partial_x^2 F(x) - \lambda  F^3(x)  = - J \delta(x)\,,
    \label{wueq}
\end{gather}
which has the following simple solution for $J>0$,
\ie 
F(x)= \sqrt{2 \over \lambda } {1
\over |x|+a}\,, \quad  a={2^{3\over 4} \over  \lambda^{1\over 4}   \sqrt{|J|}}\,,
\fe
and $F(x) \to -F(x)$ for $J<0$.
Note that the shift $a$ is positive (for the solution to make sense) and determined by matching the strength of the singularity from both sides of \eqref{wueq}.

Coming back to the interacting case, the equation we want to solve is 
\ie 
\sqrt{\lambda_{\rm T}} \left(\frac{4\Gamma\left(1-\frac{d}{4}\right)}{ d(4\pi)^d\Gamma\left(\frac{3d}{4}\right)}\right)^\frac14 (-\partial_{x_\perp}^2)^\frac{d}{4} F + \lambda_{\rm T} F^3 = J \delta(x_\perp)\,.
\label{SDEcodimone}
\fe
A similar analysis as in the classical case above gives the solution for $J>0$,
\ie 
F(x_\perp)= {C_{d,d-1}\over \lambda_{\rm T}^{1\over 4} (|x_\perp|+a(J))^{d\over 4}}\,,
\fe
where the shift can be determined by integrating \eqref{SDEcodimone} over $x$,
\ie 
a(J)= \left( {(3d-4) |J|\over 8 \lambda_{\rm T}^{1\over 4} C_{d,d-1}^3} \right)^{4\over 4-3d} \,,
\fe
which is positive for $d\geq 2$. Again for $J<0$, we have $F(x) \to -F(x)$ and the one-point function $\la \phi_{111} \ra_{\cD_{d-1}}$ is negative. This justifies the sign in \eqref{norm1PF}.

The transverse profile $F(x_\perp)$ for $\la\phi_{111}\ra_{\cD_{d-1}}$ keeps track of the defect RG flow from the trivial codimension-one defect in the UV to the nontrivial conformal defect in the IR. Here the beta-function for $J$ follows from the Callan-Symanzik equation for the one-point function, 
\ie 
\B_J=-{3d-4\over 4}J\,,
\fe
which contains a trivial fixed point at $J=0$ where the one-point function $\la\phi_{111}\ra_{\cD_{d-1}}$ vanishes, and a nontrivial strongly-coupled fixed point at $J=\infty$ (which can be seen by a redefinition in the coupling space $J \to 1/J$).

\subsection{Beta-functions and Defect RG Flows}
\label{sec:defbf}

The complete solution for $\la\phi_{111}(x)\ra_{\cD_{d-1}}$ presented in Section~\ref{sec:codimone} encodes the RG flow from the trivial to the conformal codimension-one  defects, triggered by the localized magnetic coupling in \eqref{eq:action}. One may be surprised by the simplicity of this flow since $\phi_{111}$ in very relevant on the codimension-one defect world-volume and in principle other operators including bilinears of $\phi_{abc}$ can be generated along the flow. Nevertheless, this simplification is a consequence of the large $N$ limit \eqref{lim}.
In particular, the $O(N)^3$ non-singlet operators that have higher scaling dimensions are generated but suppressed in the large $N$ limit. For instance, the operator $\phi_{111}^2$ is generated but suppressed by a factor of $\frac{1}{N^2}$. Moreover the singlet operators are also generated but do not contribute to the RG flow of the defect coupling constant $J$.
While the defect RG flow in the codimension-one case is very non-perturbative, below we study the flow of line defects in the tensor model at small $\eps=4-d$, in which case the localized magnetic coupling is weakly relevant and the flow is short. In this case, we will again derive an exact solution for the defect one-point function that encodes the entire line defect RG flow.

\subsubsection{Defect RG flow and fixed point in the tensor model}

Let us consider a general scalar QFT with a quartic potential in $d=4-\eps$ in the presence of a localized magnetic line defect. The coupled system is described by the follow action,
\begin{gather}
    S = \int d^d x  \left[\frac{1}{2} \left(\partial_\mu \phi_i\right)^2 - \frac{1}{4!} Y_{ijkl} \phi_i \phi_j \phi_k \phi_l\right] -   h  \int d\tau \phi_1(\tau,\vec{x} = 0)\,,
\end{gather}
where $Y_{ijkl}$ is a totally symmetric tensor of couplings. Using the $\eps$-expansion we can compute the two-loop contributions to
the bulk beta-function for  $Y_{ijkl}$ and up to four-loop for the defect coupling constant $h$,
\begin{gather}
    \beta_{ijkl} = -\eps Y_{ijkl} + \frac{3}{(4\pi)^2} Y_{mn(ij}Y_{kl)mn} + \frac{1}{3(4\pi)^4} Y_{mncd}Y_{d(ijk}Y_{i)mnp} - \frac{6}{(4\pi)^4} Y_{(i|mpq} Y_{|j|npq}Y_{kl)mn} \,, \notag\\
    \beta_h = - \frac{\eps}{2} h + \frac{1}{(4\pi)^2} h^3 Y_{1111} + \frac{h}{12 (4\pi)^4} Y_{1mnc}^2 + \frac{h^3}{4(4\pi)^4} Y_{11mn}^2 - \frac{h^5}{12 (4\pi)^4} Y_{111m}^2\,,
    \label{genF4beta}
\end{gather}
where the open parenthesis on the flavor indices stands for the symmetrization and   summations are implicit for all repeated indices. Now we specialize to the tensor model (i.e. with $N^3$ scalar fields), where we include all possible $O(N)^3$ invariant quartic couplings and decompose $\B_{ijkl}$ with respect to the independent invariants. The similar analysis was done for the defect-less theory in \cite{Giombi:2017dtl}. We are lead to the following action,
\begin{gather}
    S = \int d^d x \left[\frac{1}{2} (\partial_\mu \phi_{abc})^2 + \frac14 g_{\rm T} O_{\rm T} + \frac14 g_{\rm P} O_{\rm P} +\frac14 g_{\rm dt} O_{\rm dt}\right]-   h  \int d\tau \phi_1(\tau,\vec{x} = 0)\,, 
    \label{genDTM}
\end{gather}
where the quartic interactions are given by
\ie 
    O_{\rm T} =\,& \phi_{abc}\phi_{ab'c'}\phi_{a'bc'}\phi_{a'b'c}\,,\notag\\
    O_{\rm P} = \,&\frac13 \left[\phi_{abc}\phi_{abc'}\phi_{a'b'c}\phi_{a'b'c'}+\phi_{abc}\phi_{ab'c}\phi_{a'bc'}\phi_{a'b'c'}+\phi_{abc}\phi_{a'bc}\phi_{a'b'c'}\phi_{ab'c'}\right]\,, \notag\\
    O_{\rm dt} =\,& \phi_{abc}\phi_{abc}\phi_{a'b'c'}\phi_{a'b'c'}\,,
\fe 
which correspond to the tetrahedral, pillow and double-trace operators respectively.  
We then find the following system of beta-functions for the defect tensor model using \eqref{genF4beta}. 

As expected, the beta-functions for the bulk interactions coincide with those derived in \cite{Giombi:2017dtl} and are given by,
\ie 
 &\beta_{\rm T} = -\eps g_{\rm T} -\frac{\left(5 N^3+82\right) g _{\text{dt}}^2 g_{\rm T}}{64 \pi ^4}+g _{\text{dt}} \left(-\frac{(N (5
   N+17)+17) g_{\rm P} g_{\rm T}}{32 \pi ^4}-\frac{15 N g_{\rm T}^2}{32 \pi ^4}-\frac{g_{\rm P}^2}{4 \pi ^4}+\frac{3
   g_{\rm T}}{2 \pi ^2}\right)  \notag\\& +\frac{\left(N^3-15 N-10\right) g_{\rm T}^3}{128 \pi ^4}-\frac{(N (2 N+13)+24) g
   _P^3}{288 \pi ^4}+g_{\rm P}^2 \left(\frac{1}{6 \pi ^2}-\frac{(N (N (N+15)+93)+101) g_{\rm T}}{384 \pi
   ^4}\right)  \notag\\& + g_{\rm P} \left(\frac{(N+1) g_{\rm T}}{4 \pi ^2}-\frac{(N (N+4)+13) g_{\rm T}^2}{32 \pi ^4}\right)\,, \notag
   \\&
   \beta_{\rm P} =  -\eps g_{\rm P} -\frac{\left(5 N^3+82\right) g _{\text{dt}}^2 g_{\rm P}}{64 \pi ^4}+g _{\text{dt}} \left(-\frac{(N (7
   N+15)+29) g_{\rm P}^2}{32 \pi ^4}+g_{\rm P} \left(\frac{3}{2 \pi ^2}-\frac{(39 N+48) g_{\rm T}}{32 \pi
   ^4}\right) \right.\notag\\&\left.- \frac{9 (N+2) g_{\rm T}^2}{16 \pi ^4}\right) +\frac{3 (N+2) g_{\rm T}^2}{8 \pi ^2}-\frac{3 \left(N^2+N+4\right) g_{\rm T}^3}{32 \pi
   ^4}-\frac{(N (5 N (N+9)+243)+343) g_{\rm P}^3}{1152 \pi ^4}  \notag\\& + g_{\rm P}^2 \left(\frac{N (N+5)+12}{24 \pi ^2}-\frac{(2
   N (2 N+9)+29) g_{\rm T}}{32 \pi ^4}\right)
   \\
   &+g_{\rm P} \left(\frac{(N+2) g_{\rm T}}{2 \pi ^2}-\frac{(N (N
   (N+12)+99)+98) g_{\rm T}^2}{128 \pi ^4}\right)\,, \notag\\& 
\beta_{\rm dt} = -\eps g_{\rm dt} -\frac{\left(9 N^3+42\right) g _{\text{dt}}^3}{64 \pi ^4}  + g _{\text{dt}}^2 \left(-\frac{11
   \left(N^2+N+1\right) g_{\rm P}}{32 \pi ^4}+\frac{N^3+8}{8 \pi ^2}-\frac{33 N g_{\rm T}}{32 \pi ^4}\right)
   \\
   &+g
   _{\text{dt}} \left(g_{\rm P} \left(\frac{N^2+N+1}{4 \pi ^2}-\frac{(5 N (N+1)+17) g_{\rm T}}{32 \pi
   ^4}\right)  \right.\notag \left. - \frac{5 \left(N^3+3 N+2\right) g_{\rm T}^2}{128 \pi ^4} \right.
  \\&\left.-\frac{(N (5 N (N+3)+93)+97) g_{\rm P}^2}{384 \pi
   ^4}+\frac{3 N g_{\rm T}}{4 \pi ^2}\right)+g_{\rm P} \left(\frac{g_{\rm T}}{4 \pi ^2}-\frac{\left(N^2+N+4\right)
   g_{\rm T}^2}{32 \pi ^4}\right)  \notag\\& -\frac{7 (N (N+3)+5) g_{\rm P}^3}{576 \pi ^4}+g_{\rm P}^2 \left(\frac{2 N+3}{24 \pi
   ^2}-\frac{(N+1) g_{\rm T}}{8 \pi ^4}\right)-\frac{3 N g_{\rm T}^3}{64 \pi ^4}\,.
\fe 
On the other hand, the beta function for the defect coupling is
\ie 
&\beta_h = -\frac{h
   \epsilon }{2} + h^3 \left(\frac{g_{\text{dt}} \left(8 \pi ^2-\left(N^2+N+7\right) g_p\right)}{128 \pi ^4}-\frac{\left(N^3+8\right)
   g_{\text{dt}}^2}{256 \pi ^4}  g_T \left(\frac{8 \pi ^2-3 (N+2) g_p}{128 \pi ^4}-\frac{3 (N+2) g_{\text{dt}}}{128 \pi
   ^4}\right) \right.  \notag\\  & + \left. 
   +\frac{g_p \left(48 \pi ^2-(N (N+7)+19) g_p\right)}{768 \pi ^4}-\frac{3 (N+2) g_T^2}{256 \pi ^4}\right)  \notag\\ & +h^5
   \left(g_T \left(-\frac{3 g_{\text{dt}}}{128 \pi ^4}-\frac{3 g_p}{128 \pi ^4}\right)-\frac{3 g_{\text{dt}} g_p}{128 \pi
   ^4}-\frac{3 g_{\text{dt}}^2}{256 \pi ^4}-\frac{3 g_p^2}{256 \pi ^4}-\frac{3 g_T^2}{256 \pi ^4}\right) \notag\\ & +h \left(g_T
   \left(\frac{3 N g_{\text{dt}}}{128 \pi ^4}+\frac{\left(N^2+N+1\right) g_p}{128 \pi
   ^4}\right)+\frac{\left(N^2+N+1\right) g_{\text{dt}} g_p}{128 \pi ^4}+\frac{\left(N^3+2\right) g_{\text{dt}}^2}{256 \pi
   ^4} \right. \notag\\ & +\left. \frac{\left(N^3+3 N+2\right) g_T^2}{512 \pi ^4}+\frac{(N (N (N+3)+9)+5) g_p^2}{1536 \pi ^4}\right)\,.
\fe
Applying the large $N$ rescaling as in \eqref{lim}, 
\begin{gather}
    h \to J N^\frac{3}{4}, \quad g_{\rm T} \to \frac{\lambda_{\rm T}}{N^\frac32}, \quad g_{\rm P} \to \frac{\lambda_{\rm P}}{N^2}, \quad g_{\rm dt} \to \frac{\lambda_{\rm dt}}{N^3}, 
    \label{rescalecouplins}
\end{gather}
and taking the limit $N\to \infty$, we obtain 
\begin{align}
\beta_J = & -\frac{ 
   \epsilon }{2} J +\frac{J^3 \lambda _{\rm T}}{16 \pi ^2}+\frac{J \lambda _{\rm T}^2}{512 \pi ^4} -\frac{3 J^5 \lambda _{\rm T}^2}{256 \pi ^4}
   \,, \notag\\
   \beta_{\rm T} = & -\eps \lambda_{\rm T} + \frac{\lambda_{\rm T}^3}{128\pi^4}\,, \notag\\
    \beta_{\rm P} = & -\eps \lambda_{\rm P} +\frac{\lambda_{\rm P}^2}{24 \pi ^2}-\frac{\lambda_{\rm P} \lambda_{\rm T}^2}{128 \pi ^4}+\frac{3 \lambda_{\rm T}^2}{8 \pi ^2}\,
    , \notag\\
    \beta_{\rm dt} = & - \eps \lambda_{\rm dt}+\frac{\lambda _{\text{dt}}^2}{8 \pi ^2}+\frac{\lambda _{\text{dt}} \lambda_{\rm P}}{4 \pi ^2}-\frac{5 \lambda _{\text{dt}}
   \lambda_{\rm T}^2}{128 \pi ^4}+\frac{\lambda_{\rm P}^2}{12 \pi ^2}-\frac{\lambda_{\rm P} \lambda_{\rm T}^2}{32 \pi ^4}\,,
   \label{largeNbeta}
  \end{align}
which include all terms relevant for the leading order in the $\eps$-expansion. The DCFT that describes the IR limit of the localized magnetic defect \eqref{eq:action} is identified as the fixed point of these beta functions, where the couplings take the following values,\footnote{One needs to include higher order terms in $J$ and $\lambda_{\rm T}$ in \eqref{largeNbeta} in order to determine the higher order corrections  in $\epsilon$.
}
\begin{gather}
    \lambda_{\rm T} = 8\pi^2 \sqrt{2\epsilon}\,, \quad \lambda_{\rm P} = 24\pi^2\sqrt{-2 \eps}\,, \quad \lambda_{\rm dt} = - 8\pi^2 \sqrt{2} (3+\sqrt{3}) \sqrt{-\eps}\,,  \quad J=\pm \frac{\epsilon^\frac14}{2^\frac34}\,.
    \label{fpcp}
\end{gather}
The bulk fixed point couplings coincide with those found in \cite{Giombi:2017dtl} as expected. In particular, the couplings $\lambda_{\rm P}$ and $\lambda_{\rm dt}$ for the pillow and double-trace interactions are imaginary at the fixed point, which signals non-unitarity in the melonic CFT. We emphasize that since these couplings have been rescaled by large $N$ factors as in \eqref{rescalecouplins}, they are suppressed in the large $N$ limit \eqref{lim}.

\subsubsection{Exact RG for line defect at small $\eps=4-d$}
\label{sec:linedefRG}
Here we start by studying the behavior of the one-point function $\la \phi_{111}\ra_{\cD_1}$ in the region close to the line defect described by \eqref{eq:action} with $p=1$.  This corresponds to the UV limit of the defect tensor model and is controlled by a perturbation series in the defect coupling $J$. The coefficients in the $J$ expansion have specific dependence on the transverse distance $x_\perp$ that is dictated by the perturbative contributions at each order and multiplied by certain constants. By consistency, these constants are constrained by the defect Schwinger-Dyson equation \eqref{SDEdiff}. By solving these constraints, we will determine the defect one-point function along the entire defect RG flow for small $\eps=4-d$. In particular, we will see explicitly the DCFT solution found in Section~\ref{sec:confdf} emerges in the IR limit. 

For convenience, we introduce here the rescaled one-point function of $\phi_{111}$ and defect coupling $J$,
\ie 
\tilde F= \lambda_{\rm T}^{1\over 4} F \,,\quad \tilde J=\lambda_{\rm T}^{-{1\over 4}} J\,,
\fe
such that the defect Schwinger-Dyson equation \eqref{SDEdiff} can be written as
\ie 
B_d (-\Delta_\perp)^{d\over 4} \tilde F +\tilde  F^3=\tilde J  \delta^{d-1}(x_\perp)\,.
\label{SDEsimp}
\fe
We expect the following perturbative expansion for small $x_\perp$ (equivalently large $k_\perp$ in the Fourier space), 
\ie 
  \tilde F(x_\perp) = & \frac{1}{x_\perp^{\frac{d}{2}-1}}\sum^\infty_{n=0} \tilde{a}_n    \tilde J^{1+2n} \left(B_d\right)^{3n+2} x_\perp^{\left(2-\frac{d}{2}\right)n}\,, 
   \quad
   F(k_\perp) =  \frac{1}{k_\perp^\frac{d}{2}} \sum^\infty_{n=0} \frac{a_n }{k_\perp^{\left(2-\frac{d}{2}\right)n}}  \tilde J^{1+2n} \left(B_d\right)^{ 3n+2} \,,
     \label{eq:UVexpan}
\fe
where the coefficients are related by 
\ie 
\tilde{a}_n = \pi^{1-d\over 2} 2^{n(d-2)-d\over 2}  {\Gamma \left(  {n(d-4)+d-2\over 4} \right)\over \Gamma \left(  {d+n(4-d)\over 4} \right)} a_n\, .
\fe 
In particular the first coefficient $a_0$ is fixed by the source term in \eqref{SDEsimp} to be
\ie 
a_0= {1\over B_d^3}\,.
\fe
In addition, the higher order coefficients obey the following recursive relation from \eqref{SDEsimp},
\ie 
  P_{d,n}\tilde a_n = -\sum_{n_1+n_2+n_3+1=n} \tilde{a}_{n_1}\tilde{a}_{n_2}\tilde{a}_{n_3}  \,, \quad P_{d,n}={2^{d\over 2} \Gamma \left( {n(4-d)+d\over 4} \right) \Gamma \left( {n(d-4) +2(d-1)\over 4}\right)\over 
  \Gamma \left( {n(4-d)\over 4}\right)
  \Gamma \left( {n(d-4) +d-2\over 4}\right)
  } \,.
  \label{eq:RecRel}
\fe 
This equation simplifies when we expand to leading order in $\eps
=4-d$,
\ie 
\tilde a_n = -{2\over n \eps}\sum_{n_1+n_2+n_3+1=n} \tilde{a}_{n_1}\tilde{a}_{n_2}\tilde{a}_{n_3}\,,
\label{recursim}
\fe
which can be solved as follows. We introduce the generating function $\A(t)$,
\ie
\A(t)=\sum_{n=0}^\infty \tilde a_n t^n \,,
\fe
which satisfies the following differential equation due to \eqref{recursim},
\ie 
{\partial \A(t)\over \partial t}=-{2\over \eps} \A^3(t)\,.
\fe
The solution satisfying the initial condition $\A(0)=\tilde a_0$ is  given by 
\ie 
\A(t)= {\tilde a_0\over \sqrt{{4t \tilde a_0^2\over \eps} + 1}}\,.
\fe
Consequently, we have up to $\cO(\eps)$ corrections,
\ie 
\tilde F(x_\perp)= {2^{-{1\over 4}}\eps^{1\over 4} \tilde J\over x_\perp^{\frac{2-\eps}{2}}} 
{1\over \sqrt{ 8\cdot 2^{1\over 4} \pi \tilde J^2 \eps^{-{1\over 4}} x_\perp^{\eps \over 2}  +1}}\,,
\label{fullRGsmallep}
\fe
where we have used that  
\ie 
B_d={1\over 2^{7\over 3} \pi  \eps^{1\over 4}} (1+\cO(\eps))\,,\quad 
\tilde a_0=32\eps^{3\over 4} \pi^2 (1 +\cO(\eps))\,.
\fe 
Beyond small $\eps$, we can also solve the recursive relation \eqref{eq:RecRel} numerically. The numerical solutions we find for $\tilde a_n$ agree with \eqref{fullRGsmallep} for small $\eps$, providing a consistency check.

Let us now discuss some features of our solution \eqref{fullRGsmallep} which is non-perturbative in the defect coupling $\tilde J$ . We see that $\tilde J^{4/\eps}$ sets the characteristic scale of the coupled system. The solution \eqref{fullRGsmallep} has simple behaviors at small and large distance relative to this scale,
\ie 
x_\perp \ll  \tilde J^{-4/\eps}:~ \tilde F(x_\perp) \to {2^{-{1\over 4}}\eps^{1\over 4} \tilde J\over x_\perp^{\frac{d}{2}-1}}\,,
\quad 
x_\perp \gg \tilde J^{-4/\eps}:&~\tilde F(x_\perp) \to \sgn(\tilde J)\frac{\epsilon ^{3\over 8}}{  2^{15\over 8} \sqrt{\pi } x_\perp^{d\over 4} }\,.
\fe
At small distance $\tilde F(x_\perp)$ is dominated by the classical contribution with perturbative corrections in $\tilde J$, whereas at large distance it approaches the conformal solution found in Section~\ref{sec:confdf} where dependence on $\tilde J$ disappears (except for its sign).
Furthermore the solution \eqref{fullRGsmallep} captures the entire RG flow for the line defect defined as a localized magnetic perturbation in \eqref{eq:action}. It also justifies the ${\rm sgn}(J)$ factor in \eqref{norm1PF}.

\section{Defect Entropy and $g$-Function}
\label{sec:defgf}
Conformal defects of odd dimensions (i.e. $p$ odd) have a universal observable known as the \textit{defect entropy} which we denote as $s(\cD_p)$.
It is defined as the finite piece in the free energy of the defect placed on a sphere $S^p$ of size $R$,
\ie 
\la \cD_p \ra\equiv {Z_{\rm DCFT}\over Z_{\rm CFT}}\,,\quad 
\log \la \cD_p \ra = \int\limits_{S^p}  \left(   \sum_{i=0}^{(p-1)\over 2}\A_i \Lambda^{p-2i} \cR^{i}    \right) + 
s(\cD_p)\,.
\fe
Here $\cR^{i}$ denotes schematically the degree $i$ Riemann curvature invariants (which scales as ${R^{-2i}}$) and $\Lambda$ is the UV cutoff. 
While the coefficients $\A_i$ are scheme-dependent, the finite term $s(\cD_p)$ is unambiguous. For line defects, the defect entropy is related to the defect $g$-function by $s=\log g$ which was first studied in the context of conformal boundaries in $2d$ CFTs \cite{Affleck:1991tk}. 
In this case, the only possible divergence in $\log\la \cD_p\ra$ is a cosmological constant along the line and the scheme-independent defect entropy can be obtained as follows \cite{Friedan:2003yc},
\ie 
s(\cD_1)= \left (1- R{\partial \over \partial R} \right) \log\la \cD_1\ra\,,
\label{defentropy}
\fe
at large $R$.
The defect entropy (for general odd $p$) is expected to provide a measure for the degrees of freedom on the defect, playing a similar role as the sphere free energy (finite part thereof) for CFTs. In particular, the defect entropy can be defined along defect RG flows (away from the fixed points) and
has been conjectured to be monotonically decreasing under defect RG flows (see \cite{Kobayashi:2018lil} for a recent summary). This was proven recently for line defects ($p=1$) that respect reflection positivity in \cite{Cuomo:2021rkm}. While the line defect entropy is easy to determine in free theory examples, few results are known for defects in strongly coupled theories without supersymmetry. Below we will compute the line defect entropy (equivalently the $g$-function) for the localized magnetic defect defined by \eqref{eq:action} in the strongly coupled melonic CFT. We first carry out the computation exactly by summing over melonic trees with the help of the Schwinger-Dyson equations. We then calculate the defect entropy perturbatively in the $\epsilon=4-d$ expansion, providing a consistency check for our non-perturbative result and also verifying the gradient formula for defect RG flow recently proven in \cite{Cuomo:2021rkm}.

\subsection{Exact defect entropy from melonic trees}
 \label{sec:gfexact}

\begin{figure}[!htb]
    \centering
    
\begin{tikzpicture}[scale=0.75]
    \draw [line width=1.5pt] (0,0) circle (3);
    \draw[thin]  (3*cos 37,3*sin 37) node[above right] {$J$}.. controls (0,1)..(3*cos 234,3*sin 234)  node[below left] {$J$};
    \draw[thin]  (3*cos 165,3*sin 165) node[above left] {$J$}.. controls (0,1)..(3*cos 333,3*sin 333) node[below right] {$J$};
     \node[circle,fill=black,inner sep=0pt,minimum size=3pt,label=above:{$\lambda_{\rm T}$}] at (0.05,0.65) {} ;
    \begin{scope}[xshift=300]
     \draw [line width=2pt]  (0,0) circle (3);
     \draw[thin] (0,-0.35) circle (1);
      \node[circle,fill=black,inner sep=0pt,minimum size=3pt,label=above left:{$\lambda_{\rm T}$}] at (-1,-0.175) {} ;
    \node[circle,fill=black,inner sep=0pt,minimum size=3pt,label=above right:{$\lambda_{\rm T}$}] at (1,-0.42) {} ;
   \draw[thin] (3*cos 189,3*sin 189) node [left]{$J$} .. controls (-1,0) and (1,0) ..(3*cos 333,3*sin 333) node [right] {$J$};
    \end{scope}
\end{tikzpicture}
    \caption{Two infinite families of diagrams contributing to the line defect $g$-function.  Their contributions are related by a factor of two as a consequence of the Schwinger-Dyson equations.}
    \label{fig:gfun}
\end{figure}
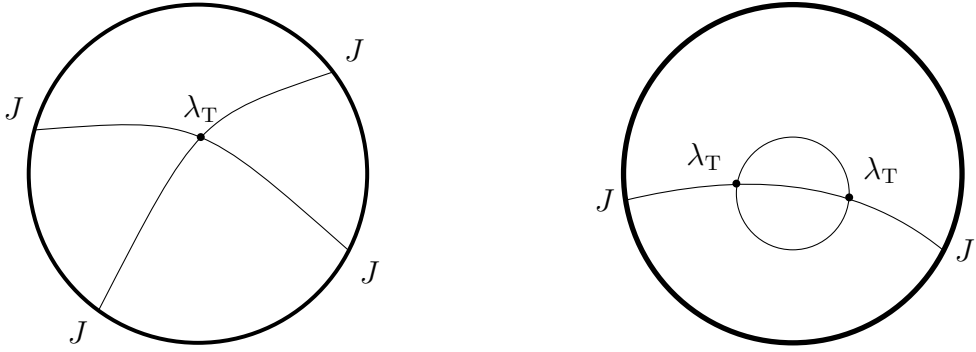

We start with the line defect stretched along the $x_2$ direction at,
\begin{gather}
    x_1 = \frac{1}{2R}\,,\quad x_3=\ldots=x_d=0\,,
\end{gather}
for which we have already determined the defect one-point function $\la \phi_{111}\ra_{\cD_1}$ in the IR DCFT (see \eqref{scant} and \eqref{norm1PF}).
By performing an inversion transformation $x_i \to \frac{x_i}{x^2}$ followed by a translation $x_1 \to x_1+R$ , we map this line to the circle located at,
\begin{gather}
    x_1^2 + x_2^2 = R^2\,, \quad x_3=\ldots=x_d=0\,.
    \label{circledef}
\end{gather}
Accordingly, the defect one-point function in \eqref{scant} transforms to,\footnote{For a review on the conformal structures of correlation functions in DCFT, we refer the readers to \cite{Billo:2016cpy}.}
\begin{gather}
  \la \phi_{111}(x)\ra = {N^{3\over 4} C_{d,1}\over \lambda_{\rm T}^{1\over 4}} \left(\frac{4 r^2}{\left(-R^2 + x_1^2 + x_2^2\right)^2 + 2 \left(R^2  + x_1^2 + x_2^2\right) x_\perp^2  + x_\perp^4}\right)^\frac{d}{8},\quad x_\perp^2 = \sum^d_{i=3} x_i^2\,.
  \label{def1PFcirc}
\end{gather}
The defect partition function $\la \cD_1\ra$ can be computed diagrammatically by summing over melonic trees (see for example Figure~\ref{fig:diags}) anchored on the defect. In this case there are two infinite families of diagrams that contribute to the defect partition function, as shown in Figure~\ref{fig:gfun}. Explicitly, the first few contributions are
\ie 
\frac{1}{N^{3\over 2}} \log \la \cD_1\ra=\,& \frac{J_0^2}{2} R^2\int d\varphi_1 d\varphi_2 \int d^d x d^d y G_0(z(\varphi_1) -z(\varphi_2))
\\
-& \frac14  \lambda_{\rm T} J_0^4  R^4\int d^d x \prod^4_{i=1}\int d\varphi_i \prod G_0\left(x-z(\varphi_i)\right)  \\ 
+ &\frac12 \lambda_{\rm T}^2 J_0^2 R^2\int d\varphi_1 d\varphi_2 \int d^d x d^d y \,G_0(z(\varphi_1) - x) G_0^3(x-y)  G_0(y-z(\varphi_2)) + \ldots\,,
\label{pertge}
\fe 
where $z(\varphi)$ parameterize the defect loop \eqref{circledef} with $\varphi\in [0,2\pi)$ as usual. Here $G_0(x)$ is the bare propagator and $J_0$ is the bare defect coupling. While the first few integrals are easy to do which we will study in Section~\ref{sec:gfgradient}, it is clear that this is not feasible to obtain results at finite (renormalized) coupling. Instead here we will take advantage the defect Schwinger-Dyson equation \eqref{SDEintegral} derived in Section~\ref{sec:defSDE} to resum all these diagrams in one shot in the large $N$ limit \eqref{lim}. 

Indeed, the contributions from the two infinite families of diagrams in Figure~\ref{fig:gfun}   have the following compact expressions,
\begin{gather}
\frac{1}{N^{3\over 2}}\log \la \cD_1\ra= -\frac14 \lambda_{\rm T}\int d^d x \braket{\phi_{111}(x)}^4_{\cD_1} + \frac12 \lambda_{\rm T}^2 \int d^d x d^d y G^3(x-y) \braket{\phi_{111}(x)}_{\cD_1}
    \braket{\phi_{111}(y)}_{\cD_1}\,.
\end{gather}
The  Schwinger-Dyson equations \eqref{eq:DSprop} and \eqref{SDEconf} further imply that the second term above is proportional to the first one by a factor of two. Therefore the total contribution can be rewritten as
\begin{gather}
\frac{1}{N^{3\over 2}} \log \la \cD_1\ra= \frac14 \lambda_{\rm T}\int d^d x \braket{\phi_{111}(x)}^4_{\cD_1}\,.
\end{gather}
To compute this integral it is convenient to work with the following ``toroidal'' coordinates,
\ie 
    x_1 =& \,R \frac{\sinh \tau}{\cosh \tau - \cos \sigma} \cos \varphi\,, \notag\\
    x_2 =& \,R \frac{\sinh \tau}{\cosh \tau - \cos \sigma} \sin \varphi
    \,, \notag\\
    x_{i+2} =&\, \hat{n}_{i} \frac{\sin \sigma}{\cosh \tau - \cos \sigma}\,,\quad i=1,2,\dots,d-2\,,
    \label{tcoord}
\fe 
where  $\hat n_i$ is a unit vector parameterizing the unit $S^{d-3}$. The ranges of these coordinates are $\tau\in [0,\infty)$, $\sigma\in [0,\pi]$ and $\varphi\in[0,2\pi]$.
The defect loop now locates at $\tau=\infty$. 
At large $\tau$ (i.e. a ``toroidal'' neighbourhood of the defect loop), the above coordinates take the following form,
\ie
    x_1 \sim R \cos\varphi + 2 R \cos \sigma \cos \varphi e^{-\tau}\,, \quad x_2 \sim R \sin\varphi + 2 R \cos \sigma \sin \varphi e^{-\tau}\,, \quad x_i \sim 2 R \sin \sigma  e^{-\tau} \hat n_i\,. \notag
\fe 
We see $\varphi$, $\sigma$ together with the unit vector $\hat  n_i$  are precisely the  coordinates on the $S^1\times S^{d-2}$ boundary of this ``toroidal'' neighbourhood.

The defect one-point function \eqref{def1PFcirc} takes a simple form in the toroidal coordinates \eqref{tcoord},
\begin{gather}
   \la  \phi_{111}(\sigma,\tau,\varphi,\vec{n}_i) \ra_{\cD_1} ={N^{3\over 4} C_{d,1}\over \lambda_{\rm T}^{1\over 4}} \frac{\left|\cos \sigma - \cosh \tau\right|^\frac{d}{4}}{R^\frac{d}{4}}\,.
\end{gather}
The line defect free energy is then given by the following integral,
\begin{gather}
\log \la \cD_1\ra = {1\over 4}\times 2\pi N^{3\over 2}  {\rm vol}_{S^{d-3}}  C_{d,1}^4 \int d\sigma d\tau \sinh\tau |\sin\sigma|^{d-3}
=   C_{d,1}^4 N^{3\over 2} { \pi^{d+1\over 2}\over \Gamma \left(d-1\over 2 \right)} \int\limits^\infty_0 d\tau \sinh \tau\,,
\label{dPFbare}
\end{gather}
where we have used that the volume form in the toroidal coordinates is
\ie 
d^d x=R^d { |\sin\sigma|^{d-3}  \sinh \tau \over |\cos\sigma-\cosh \tau|^d}\, d\sigma d\tau d\varphi  \,d{\rm vol}_{S^{d-3}}\,.
\fe
The integral \eqref{dPFbare} is UV divergent due to contributions near the defect. This can be regularized by introducing a UV cutoff $\Lambda$ that truncates the $\tau$ integral up to $ e^{\tau_*}= R \Lambda$. We then obtain at large $\Lambda$,
\begin{gather}
\log \la \cD_1\ra_{\rm reg} =   N^{3\over 2} C_{d,1}^4 { \pi^{d+1\over 2}\over \Gamma \left(d-1\over 2 \right)}
\left({1\over 2}R \Lambda - 1\right) + \mathcal{O}\left(\frac{1}{R \Lambda}\right)\,.
\end{gather}
We see the divergent term is proportional to the radius of the defect loop as expected, and the scheme-independent defect entropy can be extracted using \eqref{defentropy},
\begin{gather}
   s(\cD_1) = 
 -   N^{3\over 2} {2^{1-d}\sqrt{\pi} (d-4)^2 \Gamma \left (3d\over 4\right)^{3\over 2} \Gamma \left (1-{d\over 4}\right)^{1\over 2}  \over 
   (4-3d)^2\sqrt{d} 
   \Gamma \left(d-1\over 2 \right)\Gamma \left(d \over 4 \right)^2}\,,
   \label{deexact}
\end{gather}
which is negative for $2\leq d < 4$ and monotonically increasing from $s=-1$ at $d=2$ to $s=0$ at $d=4$. Note the curious fractional $N^{3\over 2}$ scaling for the defect entropy.

Now let us recall that the defect tensor model \eqref{eq:action} at $p=1$ describes an RG flow from the trivial line defect to the nontrivial conformal line. The trivial line obviously has zero defect entropy. Therefore this flow obeys the line defect $g$-theorem of \cite{Friedan:2003yc,Cuomo:2021rkm}.

\subsection{Defect entropy from $\eps$-expansion and gradient formula}
 \label{sec:gfgradient}

As a consistency check, we also determine the defect entropy in the $\eps=4-d$ expansion.\footnote{See \cite{Cuomo:2021kfm}  for similar computations in the $O(N)$ vector model.} We focus on the leading answer at small $\eps$ and large $N$, which receives contributions up to two-loops, as in \eqref{pertge},\footnote{The higher order terms in the couplings will contribute corrections at higher orders in $\eps$ to the defect free energy. This can be inferred from the counting $J\sim \eps^{1\over 4}$ and $\lambda_{\rm T}\sim \eps^{1\over 2}$.}
\ie 
{1\over  N^{3\over 2} }\log \la \cD_1\ra ={J_0^2\over 2} I_1 - { J_0^4 \lambda_{\rm T}\over 4} I_2 +{J_0^2\lambda_{\rm T}^2\over 2} I_3 +\cO(J_0^6 \lambda_{\rm T}^2,J_0^2 \lambda_{\rm T}^4) \,,
\label{pertg}
\fe
where $J_0$ is the bare defect coupling. The first term in \eqref{pertg} is the contribution from a bulk propagator anchored on the defect loop, while the second and third terms correspond to the diagrams in Figure~\ref{fig:gfun} at leading order in the perturbative expansion at small $\epsilon$. Explicitly, their contributions are captured by the following integrals,
\ie 
I_1=\,& R^2\int_0^{2\pi} d\varphi_1 \int_0^{2\pi} d\varphi_2 \,G_0(z(\varphi_1)-z(\varphi_2))\,,~~
I_2=R^4 \int  d^d x  \prod_{i=1}^4 \left(  \int_0^{2\pi} d\varphi_i  G_0(x-z(\varphi_i))\right)\,,
\\
  I_3 =\,&R^2\int d^d x d^d y  \int_0^{2\pi} d\varphi_1  \int_0^{2\pi} d\varphi_2 \, G_0( x-z(\varphi_1)) G_0^3(x-y) G_0(y-z(\varphi_2) )\,,
  \fe 
with the bare propagator 
\ie 
G_0(x)={f(d,2)\over (2\pi)^d |x|^{d-2}}\,.
\label{bareprop}
\fe
where the coefficient $f(d,a)$ is defined in \eqref{eq:VeryUsefulFormula}.

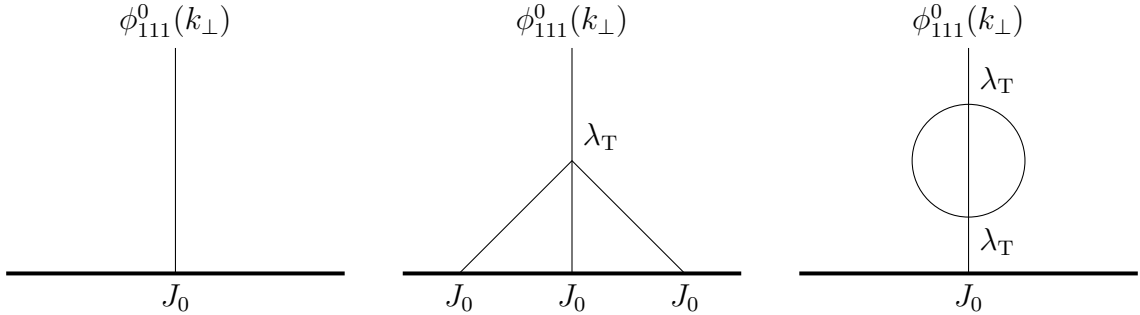
\begin{figure}[!htb]
    \centering
    \begin{tikzpicture}[scale=0.75]
        \draw[line width=1.5pt] (-3,0) -- (3,0);
        \draw (-2,0) -- (0,2)--(2,0);
        \draw (0,0) -- (0,4);
        
        \node[above right] at (0,2) {$\lambda_{\rm T}$};
        \node[above] at (0,4)  {$\phi^0_{111}(k_\perp)$};
        \node[below] at (-2,0)  {$J_0$};
        \node[below] at (0,0)  {$J_0$};
        \node[below] at (2,0)  {$J_0$};
        
        \begin{scope}[xshift=200]
        \draw (0,0) -- (0,4);
        \draw  (0,2) circle (1);
        \node[below right] at (0,1) {$\lambda_{\rm T}$};
        \node[above right ] at (0,3) {$\lambda_{\rm T}$};
        \draw[line width=1.5pt] (-3,0) -- (3,0);
        \node[above] at (0,4)  {$\phi^0_{111}(k_\perp)$};
        \node[below] at (0,0)  {$J_0$};
        \end{scope}
        
        \begin{scope}[xshift=-200]
        \draw (0,0) -- (0,4);
        \draw[line width=1.5pt] (-3,0) -- (3,0);
        \node[above] at (0,4)  {$\phi^0_{111}(k_\perp)$};
        \node[below] at (0,0)  {$J_0$};
        \end{scope}
    \end{tikzpicture}
    \caption{The diagrams that contribute to the counterterms for the defect coupling constant.}
    \label{fig:ct23}
\end{figure}

It is straightforward to evaluate the above integrals in dimensional regularization, which gives\footnote{The integrals for $I_1$ and $I_2$ can also be found in \cite{Cuomo:2021rkm}.}
\ie 
I_1=-{\eps \over 4}+\cO(\eps^2)\,,\quad 
I_2=-{1\over 16\pi^2}+ \cO(\eps) \,,\quad 
I_3={3\over 2048\pi^4}+ \cO(\eps) \,.
\label{I123v}
\fe 
For example, we perform the third integral here explicitly using  \eqref{eq:VeryUsefulFormula} and the following integration identity,
\begin{gather}
R^2\int \frac{d\varphi_1 d\varphi_2}{\left(z(\varphi_1) - z(\varphi_2)\right)^{\alpha}} = R^{2-\alpha} \frac{\pi^\frac32 2^{2-\alpha } \Gamma \left(\frac{1}{2}-\frac{\alpha }{2}\right)}{\Gamma
   \left(1-\frac{\alpha }{2}\right)}\,,
\end{gather}
which gives
\ie 
    I_3 =\, &R^2 \int \frac{d\varphi_1 d\varphi_2}{\left(z(\varphi_1) - z(\varphi_2)\right)^{3d-10}} \left(\frac{f(d,2)}{(2\pi)^d}\right)^3 f(d,3d-6) \frac{f(d,10-2d)}{(2\pi)^d}  
={3\over 2048\pi^4} + \mathcal{O}(\eps)\,.
\label{eq:I3}
\fe 
Next we need to take into account the renormalization of the defect coupling. As usual with perturbative renormalization, we split the bare coupling $J_0$ into the renormalized coupling $J$ and counterterm coefficients,
\begin{gather}
    J_0=J+ \delta J\,, \quad \delta J = \delta_2 J + \delta_3 J\,.
    \label{renormJ}
\end{gather}
These counterterms can be extracted from the diagrams depicted in Figure~\ref{fig:ct23}.
We first compute the one-point function of the bare operator $\phi^0_{111}$,
\ie 
\braket{\phi^0_{111}(k_\perp)}_{\cD_1}
=\frac{J_0}{k_\perp^2} +  \frac{\lambda_{\rm T } J_0^3}{k_\perp^{10-2d}}  \left(\frac{f(d-1,2)}{(2\pi)^{d-1}}\right)^3 f(d-1,3d-9) +    \frac{\lambda^2_{\rm T} J_0}{k_\perp^{10-2d}} \left(\frac{f(d,2)}{(2\pi)^d}\right)^3 f(d,3d-6)   \label{eq:ct3}\,.
\fe 
We take into account the renormalization $\phi_{111}^0=Z_\phi \phi_{111}$ by studying the two-point function in the bulk,
\begin{gather}
    \braket{\phi^0_{111}(k) \phi^0_{111}(-k)} = \frac{1}{k^2} + \frac{\lambda_{\rm T}^2}{k^{10-2d}} \left(\frac{f(d,2)}{(2\pi)^d}\right)^3 f(d,3d-6),
\end{gather}
and find to this order,
\ie 
Z_\phi=1-{\lambda_{\rm T}^2\over 1024\pi^4\eps}\,.
\fe
Then from the one-point function of the renormalized field $\phi_{111}$, we arrive at the following counter-term coefficients,
\begin{gather}
    \delta J = \frac{\lambda_{ \rm T} J_0^3}{32\pi^2 \eps} + \frac{\lambda^2_{ \rm T} J_0}{1024\pi^4 \eps}\,.
    \label{countertermJ}
\end{gather}
Putting together \eqref{pertg}, \eqref{I123v}, \eqref{renormJ} and \eqref{countertermJ}, we obtain
\ie 
{1\over N^{3\over 2}}\log \la \cD_1\ra 
=\,& -{J_0^2\over 8} \eps  + { J_0^4 \lambda_{\rm T}\over 64\pi^2}   +{3J_0^2\lambda_{\rm T}^2\over 4096\pi^4} +\cO(J_0^6 \lambda_{\rm T}^2,J_0^2 \lambda_{\rm T}^4)\,,
\\
=\,& -{J^2\over 8} \eps  + { J^4 \lambda_{\rm T}\over 128\pi^2}   +{J^2\lambda_{\rm T}^2\over 2048\pi^4} +\cO(J^6 \lambda_{\rm T}^2,J^2 \lambda_{\rm T}^4)\,.
\label{fEflow}
\fe
Now plugging in the fixed point values for $\lambda_{\rm T}$ and $J$ from \eqref{fpcp}, we find the defect entropy for the IR DCFT,
\ie 
s(\cD_1)\overset{\text{f.p.}}{=} \log \la \cD_1\ra \overset{\text{f.p.}}{=} - N^\frac{3}{2} \frac{\eps^\frac32}{64 \sqrt{2}}+ \cO(\eps^{5\over 2}) \,.
\fe
This matches with \eqref{deexact} in the small $\eps$ limit.

It is recently proven in \cite{Cuomo:2021rkm} that the RG flow of a line defect loop of radius $R$ satisfy a gradient formula (generalizing \cite{Friedan:2003yc}),
\ie 
R{\partial\over \partial R}  s(\cD_1)=- { R^2\over2}\int_0^{2\pi} d\varphi_1 \int_0^{2\pi} d\varphi_2 \la T_\cD(\varphi_1)T_\cD(\varphi_2)\ra_c |z(\varphi_1)-z(\varphi_2)|^2\,,
\label{gfgen}
\fe
which states the change of the defect entropy in $R$ is controlled by the connected two-point function of the defect stress tensor $T_\cD$. Here for the localized magnetic defect (see also \cite{Cuomo:2021kfm}), the defect stress tensor is $T_\cD(\varphi)=  N^{3\over 4}\B_J \phi(z(\varphi))$, where the beta-function for defect coupling (from \eqref{largeNbeta}) is
\ie 
\B_J=-\frac{ 
   \epsilon }{2} J +\frac{J^3 \lambda _{\rm T}}{16 \pi ^2}+\frac{J \lambda _{\rm T}^2}{512 \pi ^4}
   +\cO(J^5\lambda_{\rm T}^2,J^3\lambda_{\rm T}^3)\,.
\fe
We can check the gradient formula \eqref{gfgen} for our defect flow. Using the Callan-Symanzik equation, \ie 
\left( R{\partial\over \partial R} + \B_J {\partial \over \partial J}\right) \log \la \cD_1\ra  =0\,,
\fe
it follows that the gradient formula \eqref{gfgen} becomes at this order in the $\eps$-expansion,
\ie 
{\partial \over \partial J} \log \la \cD_1\ra  = {N^{3\over 2}\over 2} \B_J\,,
\fe
which is indeed satisfied by the second line of \eqref{fEflow}.

\section{Towards Defect Two-point Functions}
\label{sec:higherpoint}

Two-point functions of bulk operators in DCFT are  important observables that connect bulk and defect data by bootstrap equations \cite{Billo:2016cpy,Herzog:2020bqw}. The two-point function $\la \cO_1(x)\cO_2(y)\ra_{\cD_p}$ in the presence of a defect $\cD_p$ is no longer fixed (up to a constant) by the conformal symmetry alone. Rather it depends nontrivially on two conformally invariant cross-ratios $\xi_1$ and $\xi_2$,
\ie 
\xi_1={|x-y|^2\over 4|x_\perp||y_\perp|}\,,\quad 
\xi_2={ x_\perp \cdot y_\perp \over |x_\perp||y_\perp|}\,.
\fe
Note that $\xi_2$ is trivial for $p=d-1$.
There are two OPE channels for $\la \cO_1(x)\cO_2(y)\ra_{\cD_p}$. In the bulk channel, we take OPE of the bulk operators and decompose the two-point function into a sum over one-point functions multiplied by the bulk three-point OPE coefficients. In the defect OPE channel, we instead expand the bulk operators into local operators on the defect multiplied by bulk-defect OPE coefficients and then sum over the exchanged defect local operators. Associativity of the DCFT operator algebra demands that the two OPE decompositions must produce the same two-point function, which implies nontrivial constraints between local operator data on the defect worldvolume and OPE data in the bulk CFT.

It is generally very difficult to obtain defect two-point functions in strongly coupled CFTs which exhibit the above properties. In this section, we will explain how to obtain such two-point functions for the primary operators $\phi_{abc}$ in the defect tensor model. Furthermore we discuss how to extract one-point functions of bilinear operators in $\phi_{abc}$ from expanding the two-point function in the bulk OPE channel. 

\subsection{Two-point functions of $\phi_{abc}$}

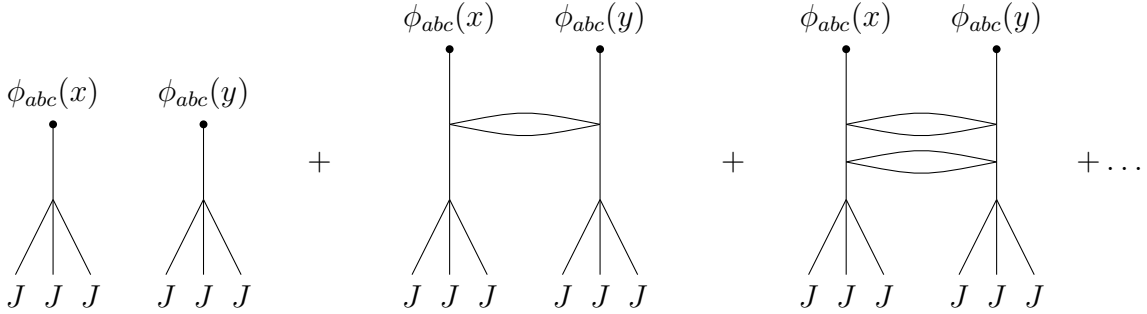
\begin{figure}[!htb]
    \centering
    \begin{tikzpicture}
        \node at (-2.5,1.5) {$+$};
        \node at (-8,1.5) {$+$};
        \node at (2.5,1.5) {$+\ldots$};
        \draw (-1,2)--(-1,3)  node[circle,fill=black,inner sep=0pt,minimum size=3pt,label=above:{$\phi_{abc}(x)$}] {} ;
        \draw (1,2)--(1,3)  node[circle,fill=black,inner sep=0pt,minimum size=3pt,label=above:{$\phi_{abc}(y)$}] {} ;
        \draw (-1,2).. controls (0,2.2) .. (1,2);
        \draw (-1,2).. controls (0,1.8) .. (1,2);
        \draw (-1,1.5).. controls (0,1.7) .. (1,1.5);
        \draw (-1,1.5).. controls (0,1.3) .. (1,1.5);
        \draw (-1,1)--(-1,2);
        \draw (1,1)--(1,2);
        \draw (-1.5,0)--(-1,1);
        \draw (-1,0)--(-1,1);
        \draw (-0.5,0)--(-1,1);
        \draw (1.5,0)--(1,1);
        \draw (1,0)--(1,1);
        \draw (0.5,0)--(1,1);
        \node[below] at (-1.5,0) {$J$};
        \node[below] at (-1,0) {$J$};
        \node[below] at (-0.5,0) {$J$};
        \node[below] at (1.5,0) {$J$};
        \node[below] at (1,0) {$J$};
        \node[below] at (0.5,0) {$J$};
        
        \begin{scope}[xshift=-150]
         \draw (-1,2)--(-1,3)  node[circle,fill=black,inner sep=0pt,minimum size=3pt,label=above:{$\phi_{abc}(x)$}] {} ;
        \draw (1,2)--(1,3) node[circle,fill=black,inner sep=0pt,minimum size=3pt,label=above:{$\phi_{abc}(y)$}] {} ;
        \draw (-1,2).. controls (0,2.2) .. (1,2);
        \draw (-1,2).. controls (0,1.8) .. (1,2);
        \draw (-1,1)--(-1,2);
        \draw (1,1)--(1,2);
        \draw (-1.5,0)--(-1,1);
        \draw (-1,0)--(-1,1);
        \draw (-0.5,0)--(-1,1);
        \draw (1.5,0)--(1,1);
        \draw (1,0)--(1,1);
        \draw (0.5,0)--(1,1);
        \node[below] at (-1.5,0) {$J$};
        \node[below] at (-1,0) {$J$};
        \node[below] at (-0.5,0) {$J$};
        \node[below] at (1.5,0) {$J$};
        \node[below] at (1,0) {$J$};
        \node[below] at (0.5,0) {$J$};
        \end{scope}
        
        \begin{scope}[xshift=-300]
         \draw (-1,1)--(-1,2) node[circle,fill=black,inner sep=0pt,minimum size=3pt,label=above:{$\phi_{abc}(x)$}] {} ;
        \draw (1,1)--(1,2) node[circle,fill=black,inner sep=0pt,minimum size=3pt,label=above:{$\phi_{abc}(y)$}] {} ;
        \draw (-1.5,0)--(-1,1);
        \draw (-1,0)--(-1,1);
        \draw (-0.5,0)--(-1,1);
        \draw (1.5,0)--(1,1);
        \draw (1,0)--(1,1);
        \draw (0.5,0)--(1,1);
        \node[below] at (-1.5,0) {$J$};
        \node[below] at (-1,0) {$J$};
        \node[below] at (-0.5,0) {$J$};
        \node[below] at (1.5,0) {$J$};
        \node[below] at (1,0) {$J$};
        \node[below] at (0.5,0) {$J$};
        \end{scope}
    \end{tikzpicture}
    \caption{The diagrams that contribute to the non-vanishing defect two-point function $\braket{\phi_{abc}(x) \phi_{abc}(y)}_{\cD_p}$ in the large $N$ limit. In comparison to case of the one-point function $\braket{\phi_{111}}_{\cD_p}$, here we have the ladder diagrams in addition to the melonic trees.}
    \label{fig:twopt0}
\end{figure}

Having understood how the non-vanishing one-point functions $\la\phi_{abc}\ra_{\cD_p}$ arises through the melonic tree diagrams in the defect tensor model, here we study the $O(N)^3$ singlet two-point function 
\ie 
 \braket{\phi_{abc}(x) \phi_{abc}(y)}_{\cD_p} \equiv N^\frac32 F_2(x,y)\,,
 \label{2pf}
\fe
by the same diagrammatic method. 
In this case in addition to the melonic tree diagrams in Figure~\ref{fig:diags}, there are other contributions of the same order in $N$ from the \textit{ladders}  diagrams in Figure~\ref{fig:twopt0}.\footnote{We emphasize that for general two-point functions of $\phi_{abc}$ that are not $O(N)^3$ singlets, the ladder contributions will be suppressed by ${1\over N^2}$ due to non-planarity.}
While the tree contributions to \eqref{2pf} are rather simple, and accounted for by the product of two one-point functions $\la \phi_{111}\ra_{\cD_p}$,  the ladder part is much more complicated. We note that the problem of the resummation of ladder contributions also arises when computing four-point functions in the tensor models or the Sachdev-Ye-Kitaev (SYK) model \cite{gross2017all,maldacena2016remarks}.

The resummation of ladders for the two-point function \eqref{2pf} can be performed using the following kernel,  
\begin{gather}
     K(x_1,x_2; x_3,x_4) = 3 \lambda_{\rm T}^2 G(x_{13}) G(x_{24}) G(x_{34})^2 = \frac{(4\pi)^d d \Gamma\left(\frac{3d}{4}\right)}{4\Gamma\left(1-\frac{d}{4}\right)} \frac{1}{|x_{13}|^\frac{d}{2}|x_{24}|^\frac{d}{2} |x_{13}|^d}\,.
\end{gather}
Consequently, the two-point function takes the following form,
\begin{gather}
    F_2(x_1,x_2) = F(x_1) F(x_2) + \int d^d x_3 d^d x_4 K(x_1, x_2; x_3, x_4) F(x_3) F(x_4) + \notag\\+ \int d^d x_3 d^d x_4 d^d x_5 d^d x_6 K(x_1, x_2; x_3, x_4) K(x_3,x_4; x_5, x_6) F(x_5) F(x_6) + \ldots\,,
\end{gather}
where $F(x)$ is defined in \eqref{Fdef}.
Formally these nested integrals represent a geometric progression and can be resummed as
\begin{gather}
    F_2(x_1,x_2) = \int d^d x_3 d^d x_4 \mathcal{F}_s(x_1,x_2;x_3,x_4) F(x_3) F(x_4)\,,\quad \mathcal{F}_s = \frac{1}{1-K}\,.
\end{gather}
We note that $\mathcal{F}_s(x_1,x_2;x_3,x_4)$ is related to the four-point functions in the (defect-less) melonic CFT \cite{maldacena2016remarks,klebanov2018tasi,liu2019d}, 
\begin{gather*}
    \braket{\phi_{abc}(x_1)\phi_{abc}(x_2)\phi_{111}(x_3)\phi_{111}(x_4)}_{\rm bulk} = \int d^d x_5 d^d x_6 \mathcal{F}_s(x_1,x_2;x_5,x_6) G(x_5,x_3) G(x_6,x_4)\,.
\end{gather*}
We thus arrive at the following relation after using the Schwinger-Dyson equation \eqref{SDEintegral},
\begin{gather}
    \frac{1}{N^\frac32}\braket{\phi_{abc}(x_1)\phi_{abc}(x_2)}_{\cD_p}  \notag\\  = \frac12 \lambda_{\rm T}^2  \int d^d x_3 d^d x_4 \braket{\phi_{abc}(x_1)\phi_{abc}(x_2)\phi_{111}(x_3) \phi_{111}(x_4)}_{\rm bulk} F^3(x_3) F^3(x_4)\,. \label{eq:OPELD} 
\end{gather}
This makes explicit how defect one-point functions together with the bulk OPE (encoded by the bulk four-point function) completely determines the defect two-point function of $\phi_{abc}$.

\subsection{One-point functions of bilinear operators}
\begin{figure}[!htb]
    \centering
    \begin{tikzpicture}
        \node at (-2.5,1.5) {$+$};
        \node at (-8,1.5) {$+$};
        \node at (2.5,1.5) {$+\ldots$};
        \node[circle, fill,inner sep=0pt,minimum size=3pt, label=above:{$\phi_{abc}\partial_{\mu_1} \ldots \partial_{\mu_s}\square^h \phi_{abc}$}] at (0,3) {};
        \draw (-1,2)--(0,3)--(1,2);
        \draw (-1,2).. controls (0,2.2) .. (1,2);
        \draw (-1,2).. controls (0,1.8) .. (1,2);
        \draw (-1,1.5).. controls (0,1.7) .. (1,1.5);
        \draw (-1,1.5).. controls (0,1.3) .. (1,1.5);
        \draw (-1,1)--(-1,2);
        \draw (1,1)--(1,2);
        \draw (-1.5,0)--(-1,1);
        \draw (-1,0)--(-1,1);
        \draw (-0.5,0)--(-1,1);
        \draw (1.5,0)--(1,1);
        \draw (1,0)--(1,1);
        \draw (0.5,0)--(1,1);
        \node[below] at (-1.5,0) {$J$};
        \node[below] at (-1,0) {$J$};
        \node[below] at (-0.5,0) {$J$};
        \node[below] at (1.5,0) {$J$};
        \node[below] at (1,0) {$J$};
        \node[below] at (0.5,0) {$J$};
        
        \begin{scope}[xshift=-150]
        \node[circle, fill,inner sep=0pt,minimum size=3pt, label=above:{$\phi_{abc}\partial_{\mu_1} \ldots \partial_{\mu_s}\square^h \phi_{abc}$}] at (0,3) {};
        \draw (-1,2)--(0,3)--(1,2);
        \draw (-1,2).. controls (0,2.2) .. (1,2);
        \draw (-1,2).. controls (0,1.8) .. (1,2);
        \draw (-1,1)--(-1,2);
        \draw (1,1)--(1,2);
        \draw (-1.5,0)--(-1,1);
        \draw (-1,0)--(-1,1);
        \draw (-0.5,0)--(-1,1);
        \draw (1.5,0)--(1,1);
        \draw (1,0)--(1,1);
        \draw (0.5,0)--(1,1);
        \node[below] at (-1.5,0) {$J$};
        \node[below] at (-1,0) {$J$};
        \node[below] at (-0.5,0) {$J$};
        \node[below] at (1.5,0) {$J$};
        \node[below] at (1,0) {$J$};
        \node[below] at (0.5,0) {$J$};
        \end{scope}
        
        \begin{scope}[xshift=-300]
         \node[circle, fill,inner sep=0pt,minimum size=3pt, label=above:{$\phi_{abc}\partial_{\mu_1} \ldots \partial_{\mu_s}\square^h \phi_{abc}$}] at (0,2) {};
        \draw (-1,1)--(0,2)--(1,1);
        \draw (-1.5,0)--(-1,1);
        \draw (-1,0)--(-1,1);
        \draw (-0.5,0)--(-1,1);
        \draw (1.5,0)--(1,1);
        \draw (1,0)--(1,1);
        \draw (0.5,0)--(1,1);
        \node[below] at (-1.5,0) {$J$};
        \node[below] at (-1,0) {$J$};
        \node[below] at (-0.5,0) {$J$};
        \node[below] at (1.5,0) {$J$};
        \node[below] at (1,0) {$J$};
        \node[below] at (0.5,0) {$J$};
        \end{scope}
    \end{tikzpicture}
    \caption{The diagrams that contribute to the non-vanishing vacuum expectation value of the bilinear operator $\phi_{abc}\partial_{\mu_1} \ldots \partial_{\mu_s}\square^h \phi_{abc}$. In comparison to the operator $\phi_{111}$, here we need to include the ladder diagrams.}
    \label{fig:twopt}
\end{figure}
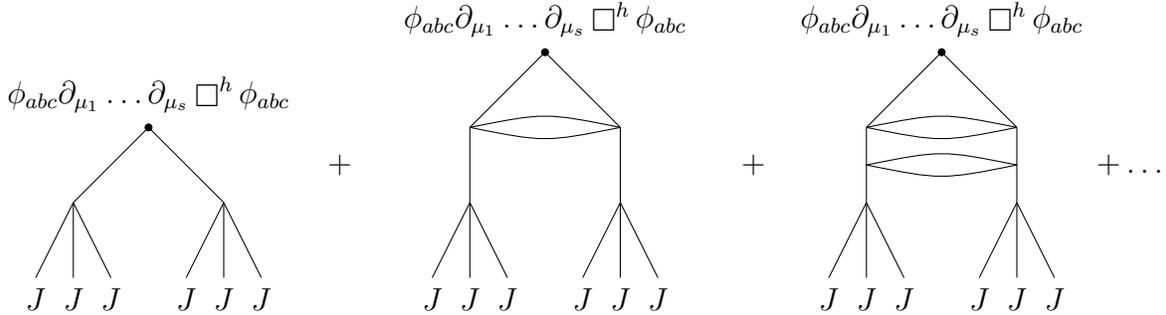

We now apply the bulk OPE in the two-point function \eqref{eq:OPELD} to compute one-point functions of the singlet bilinear operators $\phi_{abc} \partial_{\mu_1}\ldots \partial_{\mu_s} \square^h \phi_{abc}$.

In the limit $x_1 \to x_2$, we expand $\phi_{abc}(x_1)\phi_{abc}(x_2)$ into a sequence of bilinear singlet local operators. Taking this OPE limit on both sides of \eqref{eq:OPELD}, we find for a singlet operator $\cO(x)$:
\begin{gather}
    \frac{1}{N^\frac32}\braket{\cO(x)}_{\cD_p}  =\frac12  \lambda_{\rm T}^2  \int d^d x_1 d^d x_2 \braket{\cO(x)\phi_{111}(x_1) \phi_{111}(x_2)}_{\rm bulk} F^3(x_1) F^3(x_2) \label{eq:Ovev}\,,
\end{gather}
where $\cO(x)$ represents a general $O(N)^3$ singlet bilinear operator in $\phi_{abc}$.

If the operator $\cO(x)$ is a primary operator with dimension $\Delta$, we know that the three-point function is completely fixed up to an overall constant,
\begin{gather}
    \braket{\cO(x)\phi_{111}(x_1)\phi_{111}(x_2)} = \frac{C_{\cO\phi\phi}}{|x-x_1|^\Delta |x-x_2|^\Delta |x_1-x_2|^{\frac{d}{2}-\Delta}}\,,
\end{gather}
and it follows that
\begin{gather}
   \frac{1}{N^\frac32} \braket{\cO(x)}_{\cD_p} = \frac{\lambda_{\rm T}^2 C_{\cO\phi\phi} C_{d,p}^6 h_{\Delta,d}}{2|x_\perp|^\Delta}\,,
\end{gather}
where we have defined\footnote{As usual  these integrals may contain divergences that need to be regularized.}
\begin{gather}
    h_{\Delta,d} = |x_\perp|^\Delta\int \frac{d^d x_1 d^d x_2}{|x-x_1|^\Delta |x-x_2|^\Delta |x_1-x_2|^{\frac{d}{2}-\Delta} |x_{1\perp}|^\frac{3d}{4}|x_{2\perp}|^\frac{3d}{4}}\,.
\end{gather}

\subsection{One-point function of the stress-energy tensor}
Here we study the defect one-point function of the stress-energy tensor $T_{\m\n}$ in the defect tensor model.
We note that the stress tensor has in addition to a bilinear part also a quartic interaction part,
\ie 
    T_{\mu\nu} = \,&\partial_\mu \phi_{abc} \partial_\nu \phi_{abc} - \frac{1}{2} g_{\mu\nu} \left(\partial \phi_{abc} \right)^2 - \frac{d-2}{4(d-1)} \left(\partial_\mu \partial_\nu - g_{\mu\nu} \partial^2\right) \phi_{abc}^2  
    \\
    &+ \frac{\lambda_{\rm T}}{4}g_{\mu\nu} \phi_{abc} \phi_{ab'c'}\phi_{a'bc'}\phi_{a'b'c}\,.
    \fe
Following a similar analysis as in the last section, we find that the  stress tensor one-point function is given by the following expression,
\begin{gather}
    \frac{1}{N^\frac32}\braket{T_{\mu\nu}}_{\cD_p} = \frac{1}{2} \lambda_{\rm T}^2 \int d^d x_1 d^d x_2 \braket{T_{\mu\nu}(y) \phi_{111}(x_1) \phi_{111}(x_2)}_{\rm bulk} F^3(x_1) F^3(x_2) - \frac{\lambda_{\rm T}}{4} g_{\mu\nu} F^4(y)\,.
\end{gather}
We emphasize that the second term on the RHS is crucial for conformal Ward identities, ensuring $T_{\m\n}$ is traceless and conserved. Indeed the trace is given by,
\ie 
     \frac{1}{N^\frac32}\braket{T_{\mu}^\mu}_{\cD_p} =\,& \frac{1}{2} \lambda_{\rm T}^2\int d^d x_1 d^d x_2 \braket{T_{\mu}^{\mu}(y) \phi_{111}(x_1) \phi_{111}(x_2)}_{\rm bulk} F^3(x_1) F^3(x_2) - \frac{d \lambda_{\rm T}}{4}  F^4(y) \\
     =\,
    & \frac{d}{4} \lambda_{\rm T}^2 \int d^d x_1 d^d x_2 G(x_1-x_2) F^3(x_1) F^3(x_2)  - \frac{d \lambda_{\rm T}}{4}  F^4(y) = 0,
\fe 
where in the last line we have used the Schwinger-Dyson equation \eqref{SDEintegral}. More explicitly, we can write the stress tensor one-point function as,
\begin{gather}
    \frac{1}{N^\frac32} \braket{T_{\mu\nu}}_{\cD_p} = \frac{d^2 A_d C_{d,p}^6}{8(d-1)} \int d^d x_1 d^d x_2 \frac{\left(\hat{X}^{12}_\mu \hat{X}^{12}_\nu - \frac{1}{d} g_{\mu\nu}\right) |x_1-x_2|^{\frac{d}{2}}}{|y-x_2|^d |y-x_1|^d |x_{1\perp}|^\frac{3d}{4}|x_{2\perp}|^\frac{3d}{4} } \,, \notag\\
    X^{12}_\mu = \frac{(y-x_1)_\mu}{(y-x_1)^2} - \frac{(y-x_2)_\mu}{(y-x_2)^2}\,, \quad \hat{X}^{12}_\mu = \frac{X^{12}_\mu}{ |X^{12}|}\,,
    \label{T1PF}
\end{gather}
where the integrals are taken in the sense of principal value (regularization by cutting-off small spheres around $y$). One can check explicitly that \eqref{T1PF} is consistent with the Ward identities for the stress energy tensor.  Due to the residual conformal symmetry in the DCFT, the integral in \eqref{T1PF} is fixed to be of the following  universal form \cite{Billo:2016cpy},
\ie 
\la T_{\A\B}(y)\ra_{\cD_p}=-{h_p \delta_{\A\B}\over   |y_\perp|^d}\,,~ \la T_{ij}(y)\ra_{\cD_p}={(p+1)h_p    \over   (d-p-1) |y_\perp|^d} \left ( \delta_{ij} -{d\over p+1} {(y_\perp)_i (y_\perp)_j\over |y_\perp|^2}\right)\,,~ \la T_{\A i}\ra_{\cD_p}=0\,, \label{Tintegral} 
\fe
for $p\neq d-1$ otherwise $\la T_{\m\n}\ra_{\cD_p}$ vanishes identically.\footnote{While it is difficult to directly evaluate the integral \eqref{T1PF} partly due to the regularization, one can check that $\partial_{y_\perp} \la T_{dd}(y)\ra_{\cD_{d-1}}=0$ when $y_\perp \neq 0$ for the codimension-one defect. With scaling symmetry this implies that $h_{d-1}= 0$.} More explicitly, the coefficient $h_p$ can be extracted by
\begin{gather}
     h_p = N^{3\over 2}\frac{A_d C_{d,p}^6d^2 |y_\perp|^d}{8(d-1)} \int d^d x_1 d^d x_2 \frac{\hat{X}^{12}_1\hat{X}^{12}_2  |x_1-x_2|^{\frac{d}{2}}}{|y-x_2|^d |y-x_1|^d |x_{1\perp}|^\frac{3d}{4}|x_{2\perp}|^\frac{3d}{4} }\,.
\end{gather}

\section{Discussions}
 
 In the main text, we have demonstrated that the melonic tensor model provides a promising arena to study non-perturbative defects in strongly coupled CFTs. In particular, since the melonic CFTs originate from elementary scalar QFTs, one can hope to compute the defect observables from diagrammatic techniques.
 To showcase its power, we have studied the localized magnetic defect \eqref{Mdef} in the tensor model. Exploiting a new large $N$ limit \eqref{lim} that generalizes the melonic limit in the presence of such defects, we identify, at the diagrammatic level, a closed defect Schwinger-Dyson equation \eqref{SDEintegral} that resums all dominant contributions to the scalar field one-point function in this large $N$ limit. Together with the bulk Schwinger-Dyson equation that resums melonic contributions to the scalar self-energy, this allows us to determine basic defect observables such as scalar field one-point functions and defect entropy exactly in the large $N$ limit. In particular, the explicit solution we find for the scalar field one-point function  captures the exact defect RG flow from the trivial transparent defect in the UV to the nontrivial conformal defect in the IR.
 We also describe how to compute the defect two-point functions, as well as one-point functions of bilinear operators and the stress-energy tensor by taking the bulk OPE limit.
Below we discuss some future directions.

In this paper we have focused on defect observables that only involve possible local operator insertions in the bulk. An immediate generalization is to consider correlation functions that also contain nontrivial local operators on the defect world-volume. There is a set of distinguished defect local operators that are associated with bulk symmetries broken by the defect insertion. This includes the displacement operators $D^i$ for each transverse direction corresponding to the broken translation symmetries \cite{Billo:2016cpy}. Here for the localized magnetic defect, the broken part of the bulk $O(N)^3$ symmetry also give rise to $3(N-1)$ ``tilt'' operators $t_{(m)}^a$ with $a=2,3,\dots,N$ and $m=1,2,3$, analogous to those studied in \cite{Padayasi:2021sik} for the critical $O(N)$ model with a symmetry breaking boundary. These defect local operators have protected spacetime quantum numbers thanks to Ward identities for bulk currents in the presence of a conformal defect. Here the displacement operators $D^i$ appear in the  divergence of the bulk stress energy tensor 
\ie 
\partial_\mu T^{\mu i}(x)=\delta^{d-p}(x_\perp) D^i(x_\parallel)\,,
\fe
and the tilt operators modify the conservation law for the bulk $O(N)^3$ currents,
\ie 
\partial_\mu j_{(m)}^{\mu [a 1]}(x)=\delta^{d-p}(x_\perp) t_{(m)}^a(x_\parallel)\,,\quad a\neq 1\,.
\fe
Consequently, $D^i$ and $t^a_{(m)}$ are both defect scalar operators (e.g. invariant under $SO(p)$) and have scaling dimensions $\Delta(D^i)=p+1$ and $\Delta(t^a_{(m)})=p$. It would be interesting to study correlation functions involving these defect local operators in our defect tensor model.

As already emphasized in the Introduction (see also  around \eqref{fpcp}), a somewhat exotic feature of the melonic CFT is its non-unitarity, which is evident from the complex scaling dimension of $\phi_{abc}^2$ and the complex fixed point couplings in \eqref{fpcp} \cite{Giombi:2017dtl}. Nevertheless the non-unitarity is rather restricted in the leading large $N$ limit. Indeed the complex couplings in \eqref{fpcp} are all suppressed by higher powers of ${1\over N}$, and $\phi_{abc}^2$ is the only singlet primary operator of complex scaling dimension in the OPE of $\phi_{abc}$ with itself \cite{Giombi:2017dtl}. This restricted non-unitarity could be related to the fact that   all defect observables that we compute explicitly here appear to respect constraints from unitarity. It would be interesting to understand whether there is a unitary subsector in the melonic DCFT in the sense of \cite{Behan:2017rca} (generalized to DCFT).

There are close cousins of the melonic tensor models that are manifestly unitary already at finite $N$. Examples include the prismatic tensor model of \cite{Giombi:2018qgp} and the supersymmetric generalization in \cite{Popov:2019nja}. We expect the diagrammatic methods developed here can be generalized to study defects in those strongly coupled theories.

Finally the tensor models with $N^3$ degrees of freedom are reminiscent of QFTs constructed in M-theory from M5-branes \cite{Klebanov:1996un}. In particular at $d=3$, it is known that a stack of $N$ M5-branes compactified on a hyperbolic three-manifold with certain topological twist produces an $\cN=2$ supersymmetric QFT that generally flows to an interacting superconformal CFT in the IR \cite{Dimofte:2011ju,Dimofte:2011py}.\footnote{See \cite{Eckhard:2018raj} for a generalized twist that constructs 3d $\cN=1$ QFTs from the three-manifold compactification.} Such an SCFT has a sphere free energy that scales as $N^3$ at large $N$ \cite{Gang:2014ema}. It would be interesting to look for a M-theory embedding of the $d=3$ tensor model (and its cousins\footnote{For example tensor models with complex scalar fields and $SU(N)^2 \times SO(N) \times U(1)$ symmetry \cite{Klebanov:2018fzb}, as well as the cases with other interactions such as the prismatic model \cite{Giombi:2018qgp}.}) including defects therein, by 
considering a non-supersymmetric three-manifold compactification. Relatedly, it is also interesting to realize the $d=3$ tensor models via (supersymmetry breaking) deformations of 3d $\cN=2$ theories constructed in \cite{Dimofte:2011ju,Dimofte:2011py}. For this purpose, the $\cN=1$ supersymmetric tensor model in \cite{Popov:2019nja} could be a natural starting point. Such an embedding of tensor models in M-theory will provide much needed insights on their holographic dual. Indeed, the AdS$_4$ duals for the 3d $\cN=2$ SCFTs from $N$ wrapped M5-branes have been identified using M-theory \cite{Gang:2014ema}. It would be very interesting to see if suitable modifications in the bulk would produce the holographic duals for the tensor models.

 \section*{Acknowledgements}

We thank  Maikel Bosschaert and M\'ark Mezei for useful discussions and correspondences.
 We are also grateful to Dario Benedetti, Simone Giombi, Igor R. Klebanov and M\'ark Mezei for helpful comments on a draft. 
Y.W. acknowledges support from New York University and the Simons Junior Faculty Fellows program from the Simons Foundation. F.K.P.
is currently a Simons Junior Fellow at New York University and supported
by a grant 855325FP from the Simons Foundation.

 \appendix

\bibliographystyle{JHEP}
\bibliography{defects}
\end{document}